\documentclass[12pt]{article}

\usepackage{amssymb,amsfonts,amsmath,bm,color,cite,multirow,scalefnt,ulem}

\usepackage{graphicx}

\begin{document}




\vskip 2cm

\begin{center}
{\Large\bf Higher Dimensional Polytopal Universe \\ in Regge Calculus}
\end{center}
\vspace*{1cm}
\begin{center}{\sc Ren Tsuda$^{1}$ and Takanori Fujiwara$^2$}
\end{center} 
\vspace*{0.2cm}
\begin{center}
{\it $^1$Student Support Center, Chiba Institute of Technology, Narashino 275-0023, Japan} \\
{\it $^2$Department of Physics, Ibaraki University, Mito 310-8512, Japan}
\end{center}

\vfill




\begin{abstract} 
Higher dimensional closed Friedmann--Lema\^itre--Robertson--Walker 
(FLRW) universe with positive cosmological constant is investigated 
by Regge calculus. A Cauchy surface of discretized FLRW 
universe is replaced by a regular polytope in accordance with the 
Collins--Williams (CW) formalism. Polytopes in an arbitrary 
dimensions can be systematically dealt with by a set of 
five integers
integrating the Schl\"afli symbol of the polytope. 
Regge action in continuum time limit is given. 
It possesses reparameterization invariance of the time variable. Variational 
principle for edge lengths and struts yields Hamiltonian 
constraint and evolution equation. They describe oscillating 
universe in dimensions larger than three. To go beyond the 
approximation by regular polytopes, we propose pseudo-regular 
polytopes with fractional Schl\"afli symbols as a substitute 
for geodesic domes in higher dimensions. We examine the 
pseudo-regular polytope model as an effective theory of Regge 
calculus for the geodesic domes. In the infinite frequency 
limit, the pseudo-regular polytope model reduces to the 
continuum FLRW universe.
\end{abstract}

\newpage

\pagestyle{plain}




\section{Introduction}

\label{sec:intro}
\setcounter{equation}{0}

Regge calculus is a coordinate free geometric formalism of gravitation 
on triangulated piecewise linear manifolds \cite{Regge:1961aa, MTW}. 
It is envisaged from classical to quantum as an approach of 
Einstein gravity to problems where analytic methods cannot be reachable. 
Though Regge theory or its evolved ones have brought considerable 
progress in our understanding of quantum gravity, in particular in 
two and three dimensions, efforts to develop the formalism are 
vigorously continued to overcome the conceptual and technical 
difficulties \cite{BOW:2018aa}. 

As in continuum gravity, Regge calculus allows exact solutions 
for systems, where the numbers of variables are largely reduced 
by some symmetry. They are expected not only to play a role of 
a test tube to examine the validity of Regge calculus but to 
expose origins of intriguing geometrical properties of gravitation 
such as dynamical behaviors of space-time and black hole singularities. 
Along this line of thought Regge calculus has been applied 
to spherically symmetric static geometries such as the Schwarzschild 
space-time \cite{Wong:1971} and the 
Friedmann--Lema\^itre--Robertson--Walker (FLRW) 
universe \cite{CW:1973aa,Brewin:1987aa,LW:2015aa, LW:2015ab, LW:2015ac}. 
Most researches assume realistic four dimensions and
application of Regge calculus to higher dimensions have not been 
targeted so far. 

In this paper we investigate vacuum solution of a discretized 
closed FLRW universe with a positive cosmological constant in 
an arbitrary dimensions via
Regge calculus. In the previous 
papers \cite{TF:2016aa, TF:2020aa} we have analyzed 
the FLRW universe in three and four dimensions within the 
framework of Collins--Williams (CW) formalism\cite{CW:1973aa}. 
It is base on $3+1$ decomposition of space-time similar to 
Arnowitt--Deser--Misner (ADM) formalism in General 
Relativity \cite{AD:1959aa, ADM:1959aa}. Three-dimensional 
spherical Cauchy surfaces are replaced by regular polytopes 
and truncated world-tubes are 
taken as the fundamental building blocks of the discretized 
FLRW universe. 
Regge calculus describes qualitative properties 
of the continuum solution during the period small enough 
compared with the characteristic time scale $\sim1/\sqrt{\Lambda}$, 
the inverse square root of the cosmological constant. 
The deviation from the continuum theory becomes apparent as time 
passes. In three dimensions the universe expands to infinity 
in a finite time, whereas it repeats expansions and contractions
periodically in four dimensions. 

In order for Regge calculus to approximate continuum theory 
quantitatively edge lengths must be sufficiently small 
compared both with the curvature radius and
$1/\sqrt{\Lambda}$. 
This cannot be satisfied for regular 
polytopes since the edge lengths and their circumradii 
are of same order,
and the minimum edge lengths are of order 
$1/\sqrt{\Lambda}$. To improve the approximation we must 
introduce nonregular polytopes with shorter edge lengths. 
A natural construction of such polytopes is geodesic dome. 
Regge calculus for them, however, becomes impractical as 
the number of cells increases. This can be bypassed by 
working with the pseudo-regular polytopes introduced 
in \cite{TF:2016aa, TF:2020aa}. 
They can be simply defined by extending the Schl\"afli 
symbol of the original regular polytope to fractional or 
noninteger one corresponding to the geodesic dome. 
We will extend the results obtained in three and four 
dimensions to arbitrary dimensions. 

This paper is organized as follows; in the next section we set up the 
regular polytopal universe by the CW formalism in arbitrary dimensions
and formulate the Regge action in the continuum time limit. In 
Sect. \ref{sec:req} we give gauge fixed Regge equations in Lorentzian 
signature. We describe the evolution of the polytopal universe in 
detail. Comparison with the continuum solutions is made. 
In Sect. \ref{sec:prpt} we consider the pseudo-regular polytope 
having a $D$-cube as the parent regular polytope and define the 
fractional Schl\"afli symbol. Taking the infinite frequency limit, 
we argue that the pseudo-regular polytope model can reproduce the 
continuum FLRW universe. Sect. \ref{sec:sum} is devoted to summary 
and discussions. In Appendix \ref{sec:cada}, we describe 
circumradii and dihedral angles of regular polytopes in 
arbitrary dimensions.  Appendices \ref{sec:dadpf} and \ref{sec:pnu} 
are to explain some technicalities.




\section{Regge action for a regular $D$-polytopal universe}

\label{sec:ra}
\setcounter{equation}{0}

In the beginning we would like to briefly summarize the FLRW universe in General Relativity.
The continuum gravitational action with a cosmological constant in $D$ dimensions is given by 
\begin{align}
  \label{eq:cEHa}
  S=\frac{1}{16\pi}\int d^Dx\sqrt{-g}(R-2\Lambda).
\end{align}
The FLRW metric
\begin{align}
  \label{eq:FLRWm}
  ds^2=-dt^2+a(t)^2\left[\frac{dr^2}{1-kr^2}+r^2 \sigma_{AB} dx^A dx^B \right]
\end{align}
is an exact solution of Einstein's field equations, where $\sigma_{AB}$ is 
the metric tensor on $(D-2)$-dimensional unit sphere. It describes 
an expanding or contracting universe of homogeneous and isotropic 
space. All the time dependence of the metric is included in 
$a\left(t\right)$, known as scale factor in cosmology.
Einstein equations for the metric (\ref{eq:FLRWm}) derive the 
Friedmann equations as differential equations of scale factor
\begin{align}
\label{eq:Feq}
&\ddot{a}= \Lambda_D a , \quad \dot{a}^2 = \Lambda_D a^2 - k ,
\end{align}
where we have introduced $\Lambda_D$ by
\begin{align}
  \label{eq:LamD}
 \Lambda_D = \frac{2\Lambda}{ \left( D-1 \right) \left( D-2 \right) }.
\end{align}
The curvature parameter $k = 1, 0, -1$ corresponds to space being 
spherical, Euclidean, or hyperbolic, respectively. The relations 
between the solutions and curvature parameter are summarized in 
Table \ref{tab:flrw} with the proviso that the behaviors of the 
universes are restricted to expanding at the beginning for the 
initial condition $a\left(0\right)=\min a\left(t\right)$. 
Note that we have assumed $a (0) = \frac{1}{ \sqrt{\Lambda_D} }$ 
for the case of $k=0$ and $ \Lambda > 0 $.

\begin{table}[t]
\begin{align*}
   \begin{array}{cccc} \hline
   & k=1 & k=0 & k=-1 \\ \hline
   \Lambda>0 & a=\frac{1}{\sqrt{\Lambda_D}}
   \cosh\left(\sqrt{\Lambda_D}t\right) 
   & a=\frac{1}{\sqrt{\Lambda_D}}\exp\left(\sqrt{\Lambda_D}t\right) 
   & a=\frac{1}{\sqrt{\Lambda_D}}\sinh\left(\sqrt{\Lambda_D}t\right) \\
   \Lambda=0 & \mbox{no solution} & a = \mbox{const.} & a=t \\
   \Lambda<0 & \mbox{no solution} & \mbox{no solution} 
   & a=\frac{1}{\sqrt{-\Lambda_D}}\sin\left(\sqrt{-\Lambda_D}t\right) \\ \hline
   \end{array}
\end{align*}
\caption{Solutions of the Friedmann equations.}
\label{tab:flrw}
\end{table}

\begin{table}[t]
\centering
  \begin{tabular}{clll}\hline 
    & Name & $\left\{p_1,p_2,p_3,\cdots,p_D,p_0\right\}$ & 
    $[D,\kappa_D,\lambda_D,\mu_D,\zeta_D]$\\ \hline
    0-polytope & Point & $\left\{2\right\}$ & $[0,3,3,3,3]$\\
    1-polytope & Line segment & $\left\{2,2\right\}$ & $[1,3,3,3,3]$ \\
    2-polytope & $n$-sided polygon & $\left\{2,n,2\right\}$ & $[2,n,3,3,3]$ \\
    \\
    \multirow{5}{*}{3-polytope} & Tetrahedron & $ \left\{2,3,3,2\right\}$
    & $[3,3,3,3,3]$ \\ 
    & Cube & $ \left\{2,4,3,2\right\}$ & $[3,4,3,3,3]$ \\ 
    & Octahedron & $ \left\{2,3,4,2\right\}$ & $[3,3,4,3,3]$ \\ 
    & Dodecahedron & $ \left\{2,5,3,2\right\}$ & $[3,5,3,3,3]$ \\ 
    & Icosahedron & $ \left\{2,3,5,2\right\}$ & $[3,3,5,3,3]$ \\
    \\
    \multirow{6}{*}{4-polytope} & 5-cell & $ \left\{2,3,3,3,2\right\}$ 
    & $[4,3,3,3,3]$\\
    & 8-cell & $ \left\{2,4,3,3,2\right\}$ & $[4,4,3,3,3]$ \\
    & 16-cell & $ \left\{2,3,3,4,2\right\}$ & $[4,3,3,4,3]$ \\
    & 24-cell & $ \left\{2,3,4,3,2\right\}$ & $[4,3,4,3,3]$ \\
    & 120-cell & $ \left\{2,5,3,3,2\right\}$ & $[4,5,3,3,3]$ \\
    & 600-cell & $ \left\{2,3,3,5,2\right\}$ & $[4,3,3,5,3]$ \\ 
    \\
    \multirow{3}{*}{\begin{tabular}{l} $n$-polytope \\ 
        $\left(n\geq 5\right)$ \end{tabular}} & $n$-simplex $\alpha_n$ & 
    $\left\{2,3^{n-1},2\right\}$ & $[n,3,3,3,3]$ \\
    & $n$-orthoplex $\beta_n$ & $\left\{2,3^{n-2},4,2\right\}$ & $[n,3,3,3,4]$ \\ 
    & $n$-cube $\gamma_n$ & $\left\{2,4,3^{n-2},2\right\}$ & $[n,4,3,3,3]$ \\
    \hline
  \end{tabular}
  \caption{Extended Schl\"afli symbols for regular polytopes. 
    The symbol $\{2,3^4,2\}$ is an abbreviation of 
  $\{2,3,3,3,3,2\}$.
  By H. M. S. Coxeter the $n$-simplex, $n$-orthoplex, and $n$-cube are labeled as $\alpha_n$, $\beta_n$, and $\gamma_n$, respectively \cite{Coxeter}.
  The parameter set $ \left[ D, \kappa_D, \lambda_D, \mu_D, \zeta_D \right] $ is another way to specify a regular $D$-polytope introduced in Sect. \ref{sec:req}.
  }
  \label{tab:ssfrpt} 
\end{table}

As preparation for the investigation of polytopal universes, 
we work in Euclidean space-time for the time being and explain an 
epitome of Regge calculus; in Regge calculus, the discrete gravitational 
action is given by the Regge action\cite{Miller:1997aa}
\begin{align}
  \label{eq:ract}
  S_{\rm Regge}=\frac{1}{8\pi}\left(\sum_{i\in\rm \{hinges\}}
    \varepsilon_iA_i-\Lambda\sum_{i\in\rm \{blocks\}} V_i\right),
\end{align}
where $A_i$ is the volume of a hinge, $\varepsilon_i$ the deficit 
angle around the hinge of volume $A_i$, and $V_i$ the volume of a building 
block of the piecewise linear manifold.
The fundamental variables in Regge calculus are the edge lengths $l_i$. 
Varying the Regge action with respect to $l_i$, we obtain the Regge 
equations
\begin{align}
  \label{eq:regeq}
  \sum_{i\in \rm \{hinges\}}\varepsilon_i\frac{\partial A_i}{\partial l_j}
  -\Lambda\sum_{i\in \rm \{ blocks \}}\frac{\partial V_i}{\partial l_j}=0.
\end{align}
Note that there is no need to carry out the variation of the deficit 
angle owing to the Schl\"afli 
identity\cite{Schlafli:1858aa,HHKL:2015aa}
\begin{align}
\sum_{i\in\rm\{hinges\}}A_i\frac{\partial\varepsilon_i}{\partial l_j}=0.
\end{align}

We now turn to polytopal universe. According to CW formalism we 
replace $(D-1)$-dimensional hyperspherical Cauchy surface in FLRW 
universe by a fixed type of regular $D$-polytope. In general a 
regular $D$-polytope for $D \geq 2$ is characterized by a set of 
$D-1$ integer parameters $ \left\{p_2,p_3,\cdots,p_D\right\}$, known 
as Schl\"afli symbol\cite{Coxeter,Hitotsumatsu}. In this paper we 
introduce $p_0=p_1=2$ to include the cases of $D=0,~1$ and write 
the Schl\"afli symbol as $\left\{p_1,p_2 , p_3\cdots,p_D,p_0\right\}$, which 
will be referred to as extended Schl\"afli symbol. 
Each regular $D$-polytope has a corresponding dual polytope represented by the extended Schl\"afli symbol in reverse order $\left\{ p_0 , p_n , p_{n-1} , \cdots , p_1 \right\}$.
Note that there are only three types of regular polytopes in dimensions larger than 
four: the $n$-simplex, $n$-orthoplex, and $n$-cube being, 
respectively, higher dimensional analogs of the tetrahedron, octahedron, and 
cube in three dimensions. In 
Table \ref{tab:ssfrpt}\cite{TF:2020aa} we summarize all  possible 
regular polytopes in arbitrary dimensions.

\begin{figure}[t]
  \centering
  \includegraphics[scale=1]{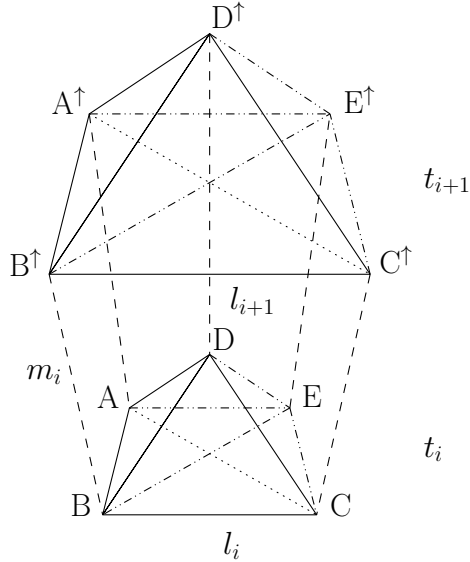}
  \caption{The $i$-th frustum as the fundamental building block 
    of the 5-polytopal universe for $\left\{2,3^4,2\right\}$.
    A lower cell like ABCDE for $ \left\{2,3,3,3,2\right\} $ with 
    edge length $l_i$ at the time $t_i$ evolves into an upper one 
    A$^\uparrow$B$^\uparrow$C$^\uparrow$D$^\uparrow$E$^\uparrow$ 
    with $l_{i+1}$ at $t_{i+1}$. The 3-frustum 
    ABC-A$^\uparrow$B$^\uparrow$C$^\uparrow$ having 2-simplices 
    $\left\{2,3,2\right\}$ as base faces is a temporal hinge, and 
    the 3-simplex ABCD for $\left\{2,3,3,2\right\}$ a spatial hinge.}
  \label{fig:fbb}
\end{figure}

In the present polytopal universe the fundamental building blocks 
of space-time are world-tubes of $D$-dimensional frustums with the 
regular $(D-1)$-polytopes $\left\{p_1,\cdots,p_{D-1},p_0\right\} $ as the 
upper and lower cells. We will refer to them as $D$-frustums. 
In Figure \ref{fig:fbb} we give, as an illustration, a depiction of 
a 5-frustum with 4-simplices as base cells. 
We assume that the 
upper and lower cells of a block lie in two consecutive time-slices 
separately and every strut between them has equal length. 
We denote the volume of the $i$-th $D$-frustum by $V_i$. It 
contains two types of the fundamental variables: the edge lengths 
$l_i$ and $l_{i+1}$ of the lower and upper $(D-1)$-polytopes, and the 
lengths of the struts $m_i$. In a $D$-dimensional piecewise linear 
manifold, hinges are ($D-2$)-dimensional objects, where curvature 
is concentrated. 
There are two types of hinges. One is temporally 
extended $(D-2)$-frustums with regular $(D-3)$-polytopes 
$\left\{p_1,\cdots,p_{D-3},p_0\right\}$ as the base cells, like the frustum 
$\mathrm{ABC\hbox{-}A}^\uparrow\mathrm{B}^\uparrow\mathrm{C}^\uparrow$ 
in Figure \ref{fig:fbb}. We call them ``temporal hinges'' and denote by 
$A_i^{({\rm t})}$ the volume of a temporal hinge between the $i$-th 
and $(i+1)$-th Cauchy surfaces. The other is spatially traversed
regular $(D-2)$-polytopes $\left\{p_1,\cdots,p_{D-2},p_0\right\}$ as 
facets of a Cauchy cell, or equivalently ridges of Cauchy surface, 
such as $\mathrm{ABCD}$. 
Note that in geometry a $(D-1)$-, $(D-2)$-, and $(D-3)$-dimensional face of 
$D$-polytope are also called a facet, ridge, and peak, respectively.  
We call the codimension two polytopes ``spatial hinges'' and denote by 
$A_i^{({\rm s})}$ the volume of the hinge lying in the $i$-th time-slice. 

We are able to write the Regge action for the polytopal universe
by counting the numbers of temporal hinges lying 
between two consecutive time-slices, spatial hinges 
in a time-slice, and $D$-frustums. They are just the numbers of peaks, ridges, and facets of the $D$-polytope, respectively. 
Let $N^{(D)}_n$ be the number of $n$-dimensional faces of 
a regular $D$-polytope, then the Regge action (\ref{eq:ract}) 
can be written as
\begin{align}
  \label{eq:regact}
  S_\mathrm{Regge}=\frac{1}{8\pi}\sum_i\left(N^{(D)}_{D-3}A^{({\rm t})}_i
    \varepsilon_i^{\rm (t)}+N^{(D)}_{D-2} A^{({\rm s})}_i\varepsilon_i^{\rm (s)}
    -N^{(D)}_{D-1}\Lambda V_i\right),
\end{align}
where $\varepsilon^{\rm ({\rm t})}_i$ and $\varepsilon_i^{\rm ({\rm s})}$ are 
the deficit angles around a temporal hinge of volume $A_i^{(\rm t)}$ and 
a spatial hinge of volume $A_i^{(\rm s)}$, respectively. The summation is 
taken over the time-slices. The volume of the frustum, those of hinges, 
and deficit angles can be expressed in terms of the fundamental 
variables $l$'s and $m$'s. 

For the purpose it is convenient to introduce the circumradius $\hat R_n$ and 
volume $\hat{\cal V}^{(n)}$ of a regular $n$-polytope 
$\Pi_n= \{p_1,p_2,\cdots,p_n,p_0\} $ with unit edge length. In 
Appendix \ref{sec:cada} we give a general formula for $\hat R_n$. 
See (\ref{eq:hatRn}) and (\ref{eq:cfsinphi}). 
The normalized 
volume $\hat{\cal V}^{(n)}$ can be obtained from the recurrence 
relation
\begin{align}
  \label{eq:rr_rpv} 
  \hat{\cal{V}}^{(n)}=\frac{N^{(n)}_{n-1}\sqrt{\hat{R}_n^2-\hat{R}_{n-1}^2}}{n} 
  \hat{\cal{V}}^{(n-1)}, \quad \hat{\cal{V}}^{(0)}=1,
\end{align}
where $\hat R_0=0$ is assumed.
It is now straightforward to write the volumes $V_i$ and 
$A_i^{(\mathrm{s,t})}$ as  
\begin{align}
  \label{eq:VAA}
   V_i &=\frac{1}{D} \hat{\cal{V}}^{(D-1)} 
   \sqrt{m_i^2-\hat{R}_{D-1}^2 \delta l_i^2}
   \frac{l^D_{i+1}-l^D_i}{l_{i+1}-l_i}, \\
   A_i^{({\rm s})}&=\hat{\cal{V}}^{(D-2)}l_i^{D-2}, \\
   A_i^{({\rm t})}&=\frac{1}{D-2}\hat{\cal{V}}^{(D-3)} 
   \sqrt{m_i^2-\hat{R}_{D-3}^2\delta l_i^2}
   \frac{l^{D-2}_{i+1}-l^{D-2}_i}{l_{i+1}-l_i},
\end{align}
where we have introduced the difference of edge 
length $ \delta l_i = l_{i+1} - l_i $.

\begin{figure}[t]
  \centering
  \includegraphics[scale=.9]{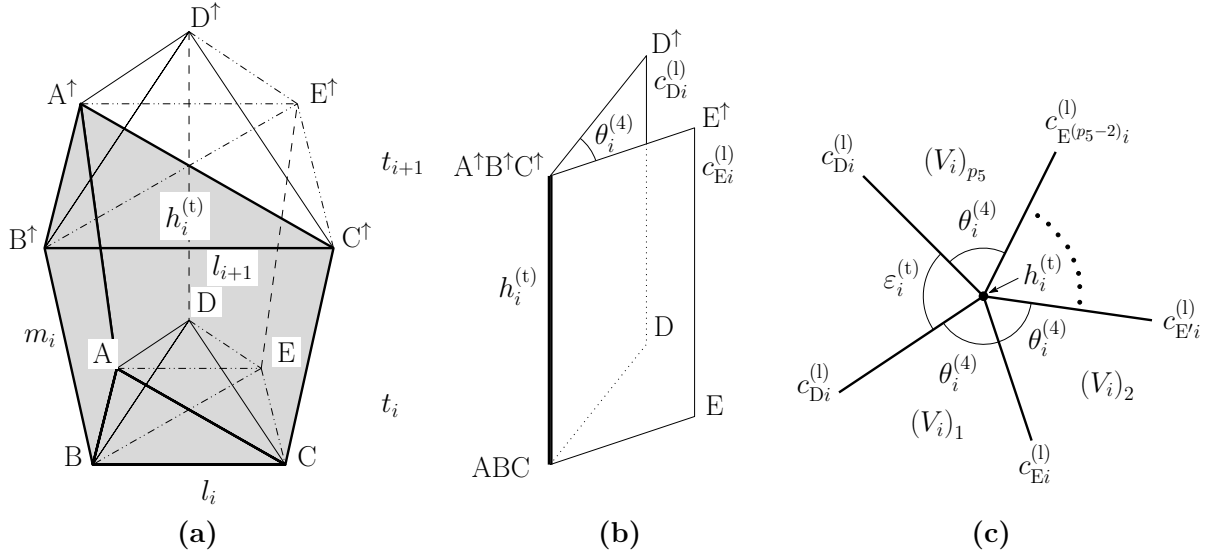}
  \caption{(a) Two lateral cells $c^{(\rm l)}_{{\rm D}i}$ and 
    $c^{(\rm l)}_{{\rm E}i}$ are meeting at a temporal hinge 
    $h^{(\rm t)}_i$, (b)  $\theta^{(4)}_i$ is the dihedral 
    angle between these cells, and (c) $\varepsilon^{(\rm t)}_i$ 
    the deficit angle around the hinge $h^{(\rm t)}_i$ made by 
    $p_5$ frustums $\left(V_i\right)_1,\cdots,\left(V_i\right)_{p_5}$ 
    having $ h^{(\rm t)}_i $ as a lateral cell in common.}
  \label{fig:tda} 
\end{figure}

\begin{figure}[t]
  \centering
  \includegraphics[scale=.9]{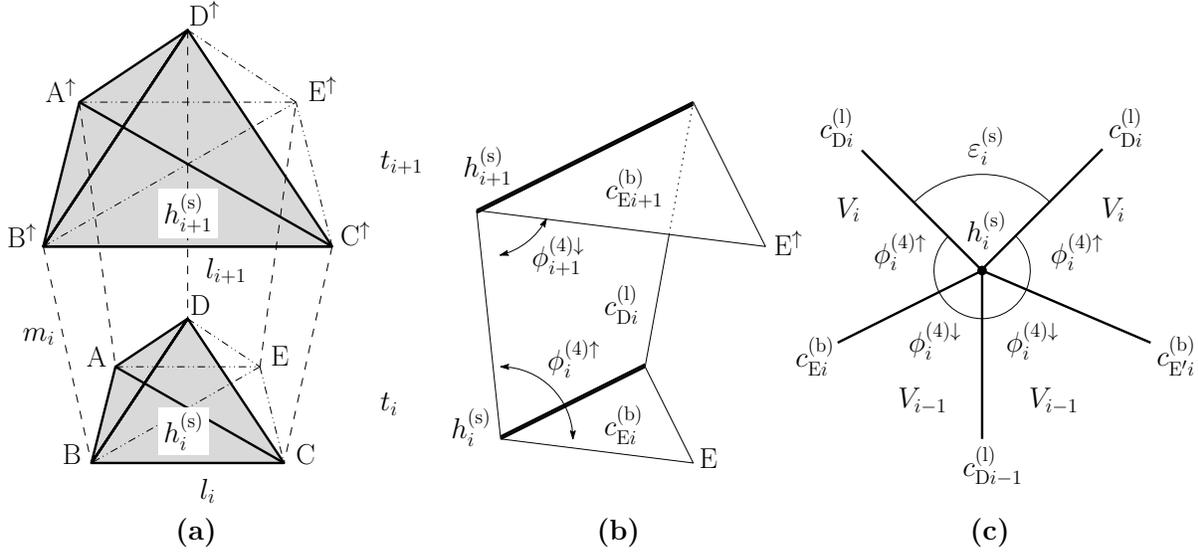}
  \caption{ (a) Two spatial hinges $h^{(\rm s)}_i$ and $h^{(\rm s)}_{i+1}$ in the $i$-th frustum,
  (b) dihedral angles $\phi^{(4)\uparrow}_i$ and $\phi^{(4)\downarrow}_{i+1}$,
  and (c) deficit angle $ \varepsilon^{(\rm s)}_i $.
    }
  \label{fig:sda} 
\end{figure}

To find the deficit angle around a hinge we need a dihedral 
angle between two adjacent cells jointed at the hinge. As an example 
consider the hinges of a 5-frustum with regular 4-polytopal bases as 
laid out in Figure \ref{fig:fbb}. At the temporal hinge 
$h^{(\rm t)}_i=\mathrm{ABC\hbox{-}A^\uparrow B^\uparrow C^\uparrow}$ in 
Figure \ref{fig:tda}, the dihedral angle $\theta^{(4)}_i$ is made by two 
lateral cells 
$c^{(\rm l)}_{{\rm D}i}=\mathrm{ABCD\hbox{-}A^\uparrow B^\uparrow C^\uparrow D^\uparrow}$ 
and $c^{(\rm l)}_{{\rm E}i} =\mathrm{ABCE\hbox{-}A^\uparrow B^\uparrow C^\uparrow E^\uparrow}$.
On the other hand, $\phi^{(4)\uparrow}_i$ is the dihedral angle at the hinge 
$h^{(\rm s)}_i =\mathrm{ABCD}$ between the lateral cell $c^{(\rm l)}_{{\rm D}i}$ 
and the lower base cell $c^{(\rm b)}_{{\rm E}i}=\mathrm{ABCDE}$ as illustrated 
in Figure \ref{fig:sda}, and similarly $\phi^{(4)\downarrow}_{i+1}$ the 
one between $c^{(\rm l)}_{{\rm D}i}$ and 
$c^{(\rm b)}_{{\rm E}i+1}=\mathrm{A^\uparrow B^\uparrow C^\uparrow D^\uparrow E^\uparrow}$ 
at $h^{(\rm s)}_{i+1} =\mathrm{A^\uparrow B^\uparrow C^\uparrow D^\uparrow }$.
For a $D$-frustum with $(D-1)$-polytopal bases, the dihedral angles 
$\theta^{(D-1)}_i$ and $\phi^{(D-1)\downarrow}_{i+1}$ can be written as
\begin{align}
   \label{eq:theta_ptu}
   \theta_i^{(D-1)} &=2\arccos\left[\sqrt{\frac{m_i^2-\hat R^2_{D-1}\delta l_i^2}{%
         m_i^2-\hat R^2_{D-2}\delta l_i^2}}\cos\frac{\vartheta_{D-1}}{2}\right], \\
   \label{eq:phi_down_ptu}
   \phi_{i+1}^{(D-1)\downarrow} &=\arccos\left[\sqrt{\frac{\hat R^2_{D-1}-\hat R_{D-2}^2}{%
         m_i^2-\hat R^2_{D-2}\delta l_i^2}}\:\delta l_i\right],
\end{align}
where $\vartheta_n$ is the dihedral angle of a regular $n$-polytope $\Pi_n$. 
Since the upper cell of the $D$-frustum is parallel to the lower, 
$\phi^{(D-1)\uparrow}_i$ and $\phi^{(D-1)\downarrow}_{i+1}$ satisfy
\begin{align}
\phi^{(D-1)\uparrow}_i + \phi^{(D-1)\downarrow}_{i+1} = \pi.
\end{align}
In Appendix \ref{sec:cada} we give a short account of dihedral angles of 
regular polytopes. Derivations of (\ref{eq:theta_ptu}) and (\ref{eq:phi_down_ptu}) 
are given in Appendix \ref{sec:dadpf}. 

Taking it into the consideration that $p_D$ 
frustums have a temporal hinge in common as in Figure \ref{fig:tda}(c), 
the deficit angle $\varepsilon_i^{({\rm t})}$ is given by
\begin{align}
  \label{eq:dati} 
  \varepsilon_i^{({\rm t})}=2\pi- p_D \theta_i^{(D-1)}.
\end{align}
On the other hand the spatial hinge $h^{(\rm s)}_i$ is always 
shared by four frustums as illustrated in Figure \ref{fig:sda}(c): 
two adjacent blocks of volume $V_i$ in the future side and two $V_{i-1}$ in the 
past side. Thus the deficit angle $\varepsilon_i^{({\rm s})}$ is 
expressed as
\begin{align}
  \label{eq:dasi} 
  \varepsilon_i^{({\rm s})}=2 \pi-2\left(\phi_i^{(D-1)\uparrow}
    +\phi_i^{(D-1)\downarrow}\right)
  =2\delta\phi_i^{(D-1)\downarrow},
\end{align}
where $\delta\phi_i^{(D-1)\downarrow}
=\phi_{i+1}^{(D-1)\downarrow}-\phi_i^{(D-1)\downarrow}$. 

A facet of regular $D$-polytope is a $(D-1)$-polytope having 
$N^{(D-1)}_{D-2}$ ridges of the $D$-polytope and a ridge is shared 
by two facets, so that $N^{(D)}_{D-1}$, $N^{(D-1)}_{D-2}$, and 
$N^{(D)}_{D-2}$ satisfy 
$N^{(D-1)}_{D-2}N^{(D)}_{D-1}=2N^{(D)}_{D-2}$. 
Likewise, a ridge has $N^{(D-2)}_{D-3}$ peaks of the $D$-polytope and a peak joints two 
ridges in a facet, so a facet has $\dfrac{N^{(D-1)}_{D-2}N^{(D-2)}_{D-3}}{2}$ 
peaks. Taking it into account of the fact that a peak connects $p_D$ 
facets, we find a relation
$\dfrac{N^{(D-1)}_{D-2} N^{(D-2)}_{D-3} }{2}N^{(D)}_{D-1}
=p_DN^{(D)}_{D-3}$. These constraints together with (\ref{eq:rr_rpv})
lead to 
\begin{align}
  \label{eq:RatioAs}
  \frac{N^{(D)}_{D-2}\hat{\cal V}^{(D-2)}}{N^{(D)}_{D-3}\hat{\cal V}^{(D-3)}}
  =&\frac{p_D}{D-2}
  \sqrt{\hat R_{D-2}^2-\hat R_{D-3}^2}, \\
  \label{eq:RatioV}
  \frac{N^{(D)}_{D-1}\hat{\cal V}^{(D-1)}}{N^{(D)}_{D-3}\hat{\cal V}^{(D-3)}} 
  =&\frac{2p_D}{(D-1)(D-2)}
  \sqrt{(\hat R_{D-1}^2-\hat R_{D-2}^2)(\hat R_{D-2}^2-\hat R_{D-3}^2)} \nonumber \\
  =&\frac{2p_D}{(D-1)(D-2)}
  (\hat R_{D-2}^2-\hat R_{D-3}^2)\tan\frac{\vartheta_{D-1}}{2},
\end{align}
which can be used to factor out the three couplings 
appearing 
in the action (\ref{eq:regact}). As for the second equality in 
(\ref{eq:RatioV}), use has been made of (\ref{eq:Rnrecr}). We thus obtain 
\begin{align}
  \label{eq:regactd}
  S_\mathrm{Regge}=&\frac{N^{(D)}_{D-3}\hat{\cal{V}}^{(D-3)}}{8\pi}\sum_i
  \Biggl(\frac{1}{D-2} 
  \sqrt{m_i^2-\hat{R}_{D-3}^2\delta l_i^2}
  \frac{l^{D-2}_{i+1}-l^{D-2}_i}{l_{i+1}-l_i}
  \varepsilon_i^{\rm (t)} \nonumber\\
  &+\frac{2p_D}{D-2}
  \sqrt{\hat R_{D-2}^2-\hat R_{D-3}^2}l_i^{D-2}
  \delta\phi_i^{(D-1)\downarrow} \nonumber\\
  &-\frac{p_D\Lambda_D}{D}(\hat R_{D-2}^2-\hat R_{D-3}^2)
  \sqrt{m_i^2-\hat{R}_{D-1}^2 \delta l_i^2}
  \frac{l^D_{i+1}-l^D_i}{l_{i+1}-l_i}\tan\frac{\vartheta_{D-1}}{2}\Biggr).
\end{align}
In later sections we are interested in the continuum time limit. 
We replace $l_i$ and $m_i$ by $l(\tau)$ and $n(\tau)\delta \tau$, where 
$\tau$ is an arbitrary parameter and $n(\tau)$ can be regarded as lapse 
function in ADM formalism. The continuum limit $\delta\tau\rightarrow d \tau$ 
of the action can easily be obtained from (\ref{eq:regactd}) as  
\begin{align}
  \label{eq:ctregact}
  S_\mathrm{Regge}=&\frac{N^{(D)}_{D-3}\hat{\cal{V}}^{(D-3)}}{8\pi}
  \int d\tau
  \Biggl( 
  \sqrt{n^2-\hat{R}_{D-3}^2\dot l^2}\:l^{D-3}
  \varepsilon^{\rm (t)}
  -2p_D
  \sqrt{\hat R_{D-2}^2-\hat R_{D-3}^2}\:l^{D-3}\dot l
  \phi^{(D-1)\downarrow} \nonumber\\
  &-p_D\Lambda_D(\hat R_{D-2}^2-\hat R_{D-3}^2)
  \sqrt{n^2-\hat{R}_{D-1}^2\dot l^2}\:l^{D-1}\tan\frac{\vartheta_{D-1}}{2}\Biggr),
\end{align}
where $\dot l=\dfrac{dl}{d\tau}$ and total $\tau$ derivative 
terms are suppressed. We have also introduced continuum limits 
of (\ref{eq:theta_ptu}), (\ref{eq:phi_down_ptu}), and (\ref{eq:dati})
by 
\begin{align}
  \label{eq:ctheta_ptu}
  \varepsilon^{(\mathrm{t})}&=2\pi-p_D\theta^{(D-1)} \quad
  \hbox{with} \quad
  \theta^{(D-1)}=2\arccos\left[\sqrt{\frac{n^2-\hat R^2_{D-1}\dot l^2}{%
         n^2-\hat R^2_{D-2}\dot l^2}}\cos\frac{\vartheta_{D-1}}{2}\right], \\
   \label{eq:cphi_down_ptu}
   \phi^{(D-1)\downarrow} &=\arccos
   \left[\sqrt{\frac{\hat R^2_{D-1}-\hat R_{D-2}^2}{%
         n^2-\hat R^2_{D-2}\dot l_i^2}}\:\dot l\right].
\end{align}
The Regge action (\ref{eq:ctregact}) is invariant under an
arbitrary reparameterization
\begin{align}
  \label{eq:rep}
  \tau \to \tau'=f(\tau), \quad
  n (\tau) \to n'(\tau')=\frac{n(\tau)}{\dot f(\tau)}, \quad
  l (\tau) \to l'(\tau')=l(\tau).
\end{align}
This can be used to fix the lapse function.




\section{Regge equations}
\label{sec:req}
\setcounter{equation}{0}

The Regge equations can be obtained by taking variations of 
the Regge action with respect to $n$ and $l$. The equations 
of motion possess the local symmetry (\ref{eq:rep}). 
We must fix it by imposing some condition on the dynamical 
variables. Furthermore, the action is based on the piecewise 
linear manifold with Euclidean signature. We must carry out 
inverse Wick rotation to recover Lorentzian signature. As for 
fixing the local invariance we impose the following gauge 
condition on the lapse function 
\begin{align}
  \label{eq:gconn}
  n(\tau)=1.
\end{align}
We then carry out inverse Wick rotation by $\tau=it$, where $t$ 
can be regarded as the time of a clock fixed 
at a vertex of the  polytopal universe. The time axis is taken 
to be parallel to a strut. It is not orthogonal to Cauchy cells. 
If we consider nonregular polytopes with shorter edge lengths 
and more cells such as geodesic domes \cite{TF:2016aa}, we 
would have a better approximation of a smooth hypersphere. 
The orthogonality of the time axis with the spatial ones as 
in the FLRW universe can be restored in the limit of smooth 
hypersphere. We thus obtain the Regge equations 
\begin{align}
  &2\pi-p_D\theta^{(D-1)}
  =p_D\Lambda_D(\hat{R}_{D-2}^2-\hat{R}_{D-3}^2)
  \sqrt{\frac{1+\hat{R}_{D-3}^2\dot{l}^2}{1+\hat{R}_{D-1}^2\dot{l}^2}}
    \:l^2\tan\frac{\vartheta_{D-1}}{2}, 
  \label{eq:chc_ptu} \\
  &\frac{\ddot l}{1+\hat R_{D-2}^2\dot l^2}
    =\Lambda_D l\left[1+\hat R_{D-3}^2\dot l^2
    -\frac{(\hat R_{D-1}^2-\hat R_{D-3}^2)l\ddot l}{%
    2(1+\hat R_{D-1}^2\dot l^2)}\right],
\label{eq:cev_ptu}
\end{align}
where the dots on $l$ stand for $t$ derivatives and 
$\theta^{(D-1)}$ in lorentzian signature is given by
\begin{align}
  \label{eq:lstheta}
  \theta^{(D-1)}=2\arccos\left[\sqrt{\frac{1+\hat R^2_{D-1}\dot l^2}{%
         1+\hat R^2_{D-2}\dot l^2}}\cos\frac{\vartheta_{D-1}}{2}\right].
\end{align}
Eq. (\ref{eq:chc_ptu}) is known as the Hamiltonian constraint in 
ADM formalism of canonical General Relativity. The equation of 
motion for $l$ is referred to as the evolution equation. 
We have simplified the evolution equation by using the Hamiltonian 
constraint. It is straightforward to show that the evolution 
equation can be obtained as the consistency of the Hamiltonian 
constraint with the time-development. We also mention that 
(\ref{eq:chc_ptu}) and (\ref{eq:cev_ptu}) reproduce the results 
of Refs. \cite{TF:2016aa,TF:2020aa} in three and four dimensions.

It is convenient to express the solution to the Regge equations 
in terms of the dihedral angle $\theta=\theta^{(D-1)}$. Solving 
(\ref{eq:chc_ptu}) and (\ref{eq:lstheta}) with respect to 
$l^2$ and $\dot l^2$, we obtain 
\begin{align}
  \label{eq:lsqr}
  l=&\sqrt{\frac{(2\pi-p_D\theta)\cot\frac{\theta}{2}}{%
    p_D\Lambda_D(\hat R_{D-2}^2-\hat R_{D-3}^2)}}, \\
  \label{eq:dsqr}
  \dot l=&\pm\frac{1}{\hat R_{D-2}}\sqrt{\frac{\cos\theta-\cos\theta_0}{%
      \cos\theta_\mathrm{c}-\cos\theta}},
\end{align}
where $\theta_0=\vartheta_{D-1}$ stands for the dihedral angle of a Cauchy cell $ \left\{ p_1 , p_2 , \cdots , p_{D-1} , p_0 \right\} $ and determines the minimum size of 
the universe.
$\theta_\mathrm{c}$ is defined by
\begin{align}
  \label{eq:jc}
  \theta_{\mathrm{c}}=2\arcsin\left[\frac{\hat R_{D-3}}{\hat R_{D-2}}
    \sin\frac{\vartheta_{D-1}}{2}\right]. 
\end{align}
The velocity $\dot l$ diverges for $\theta=\theta_\mathrm{c}$, where 
the edge length becomes maximum. In three dimensions $\theta_\mathrm{c}=0$ 
since $\hat R_0=0$. 
It matches $\vartheta_1$ the dihedral angle of a 1-polytope $ \left\{ p_1 , p_0 \right\} $.
See (\ref{eq:vth1}).
In dimensions larger than three $\theta_\mathrm{c}$  
equals a dihedral angle of a regular polytope corresponding to extended 
Schl\"afli symbol $\{p_1,p_3,\cdots,p_{D-1},p_0\}$, 
which is a vertex figure of a Cauchy cell. 
For the vertex figure, see Appendix \ref{sec:cada}.

Eliminating $l$ from (\ref{eq:lsqr}) and (\ref{eq:dsqr}),
we can derive the differential equation for $\theta$
\begin{align}
  \dot\theta&=\mp\frac{2\sqrt{p_D\Lambda_D(2\pi-p_D\theta)\sin\theta}}{%
    2\pi-p_D(\theta-\sin \theta)}
  \frac{\sin\frac{\theta}{2}}{\sin\frac{\theta_0}{2}}
  \sqrt{\frac{(\cos\theta_{\mathrm{c}}-\cos\theta_0) 
      (\cos\theta-\cos\theta_0)}
    {\cos\theta_{\mathrm{c}}-\cos\theta}}.
  \label{eq:tde_ptu} 
\end{align}
The upper sign corresponds to expanding universe and the lower 
to shrinking one. This leads to an integral 
representation
\begin{align}
  \label{eq:time}
  t \left( \theta \right) =\pm\frac{1}{2\sqrt{p_D\Lambda_D}}
  \int_\theta^{\theta_0} du\frac{2\pi-p_D(u-\sin u)}{%
    \sqrt{(2\pi-p_Du)\sin u}}\frac{\sin\frac{\theta_0}{2}}{%
    \sin\frac{u}{2}}
  \sqrt{\frac{\cos\theta_{\mathrm{c}}-\cos u}{%
      (\cos\theta_{\mathrm{c}}-\cos\theta_0)(\cos u-\cos\theta_0)}},
\end{align}
where $\theta_{\mathrm c} \leq \theta\leq\theta_0$. We have assumed the initial 
condition
\begin{align}
  \label{eq:initc}
  \theta(0)=\theta_0.
\end{align}
As a function of $t$, the dihedral angle $\theta$ 
is even 
and monotonically decreasing from $\theta_0$ to 
$\theta_\mathrm{c}$ for $0\leq t\leq\tau_\mathrm{p}/2$, where $\tau_\mathrm{p}$ is given by 
$\tau_\mathrm{p}=2t(\theta_\mathrm{c})$. We can extend $\theta(t)$ as a 
continuous periodic function for arbitrary $t$ by
\begin{align}
  \label{eq:tau}
  \theta(t+\tau_\mathrm{p})=\theta(t). 
\end{align}
The edge length (\ref{eq:lsqr}) is also a periodic function of $t$. 
It is continuous for $D\geq4$, while $l$ diverges for 
$\theta \left( \tau_\mathrm{p}/2 \right) =\theta_{\mathrm{c}}=0$ 
in three dimensions. 
Note that $\dot l/l$ not only diverges 
but also has a discontinuity at $t=\pm\tau_\mathrm{p}/2,
~\pm3\tau_\mathrm{p}/2,~\cdots$. At present it is only an 
assumption that the polytopal universe in four or more dimensions 
jumps from expansion to contraction when it reaches the maximum size.

In dimensions larger than four there are only three types of regular 
polytopes. As can easily be seen from Table \ref{tab:ssfrpt} any regular 
polytope can be characterized by $p_2$, $p_3$, $p_D$, and $D$. It is 
possible to write the circumradii $\hat R_{D-k}$ ($k=1,2,3$) and 
dihedral angles $\vartheta_{D-1}$ appearing 
in (\ref{eq:chc_ptu})--(\ref{eq:lstheta}) in more tractable forms by noting (\ref{eq:hatRD}) 
and (\ref{eq:vTD}). To this end we define a set of parameters 
$\kappa_n$, $\lambda_n$, $\mu_n$, and $\zeta_n$ by
\begin{align}
  \kappa_n &= 3 \sum_{j=0}^1\delta_{j,n}+p_2\sum_{j=2}^\infty\delta_{j,n}, \\
  \lambda_n&=3\sum_{j=0}^2\delta_{j,n}+p_3\sum_{j=3}^\infty\delta_{j,n},\\
  \mu_n&=3\sum_{j=0}^3\delta_{j,n}+p_4\sum_{j=4}^\infty\delta_{j,n},\\
  \zeta_n&=3\sum_{j=0}^4\delta_{j,n}+p_n\sum_{j=5}^\infty\delta_{j,n},
\end{align}
where $\delta_{j,k}$ is the Kronecker delta. 
Obviously, $\kappa_n=p_2$, $\lambda_n=p_3$, $\mu_n=p_4$, and $\zeta_n=p_n$ for $n\geq5$. We assign a regular 
$D$-polytope to a set of five parameters $\left[D,\kappa_D,
\lambda_D,\mu_D,\zeta_D \right]$. In Table \ref{tab:ssfrpt} we summarize the correspondence 
between regular polytopes and the symbol $\left[D,\kappa_D,
\lambda_D,\mu_D,\zeta_D\right]$. 
This allows us to express the normalized circumradius $\hat{R}_D$ and the dihedral angle $\vartheta_D$ in the closed forms as
\begin{align}
  \label{eq:hRD}
\hat{R}_D &= \frac{1}{2} \sqrt{ \frac{ \left[ 1 - \left( D-4 \right) \cos \frac{ 2 \pi }{ \zeta_D } \right] \sin^2 \frac{\pi}{\lambda_D} - 2 \left[ 1 - \left( D - 5 \right) \cos \frac{ 2 \pi }{ \zeta_D } \right] \cos^2 \frac{ \pi }{ \mu_D } }{%
\left[ 1 - \left( D-4 \right) \cos \frac{2 \pi}{\zeta_D} \right] \left( \sin^2 \frac{\pi}{\lambda_D} - \cos^2 \frac{\pi}{\kappa_D}  \right) 
- 
2 \left[ 1 - \left( D-5 \right) \cos \frac{2 \pi}{\zeta_D} \right] \sin^2 \frac{\pi}{\kappa_D} \cos^2 \frac{ \pi}{\mu_D} 
} }, \\
\label{eq:vthD}
\vartheta_D &= 2 \arcsin \left( \sqrt{ 2 \frac{ \sin^2 \frac{ \pi }{ \kappa_D } \left[ 1 - \left( D - 5 \right) \cos \frac{ 2 \pi }{ \mu_D } \right] - \left( D - 4 \right) \cos^2 \frac{\pi}{ \lambda_D } }{%
\sin^2 \frac{ \pi }{ \kappa_D } \left[ 1 - \left( D - 4 \right) \cos \frac{ 2 \pi }{ \mu_D } \right] - \left( D - 3 \right) \cos^2 \frac{\pi}{ \lambda_D }
} } \cos \frac{\pi}{ \zeta_D } \right).
\end{align}
The circumradius (\ref{eq:hRD}) is applicable in $D \geq 0$,
whereas the dihedral angle (\ref{eq:vthD}) is valid in the dimensions larger than zero.
Note that $\vartheta_0$ is undetermined.
In particular the fact that $\mu_{D-k}=\zeta_{D-k}=3$ with $ 1 \leq k \leq D $ for any regular polytope 
enables us to write the following equalities 
\begin{align}
  \label{eq:RDk}
  \hat{R}_{D-k}
   &=\displaystyle \frac{1}{2} \sqrt{%
     \frac{\left(D-1- k\right)-2\left(D-2-k\right)
       \cos^2\frac{\pi}{\lambda_{D-1}}}
     {%
      \left(D-1-k\right)\sin^2\frac{\pi}{p_2}-2\left(D-2-k\right)
      \cos^2\frac{\pi}{\lambda_{D-1}}}} \quad \left(k=1,2,3\right), \\
  \label{eq:cosvthD}
  \cos\vartheta_{D-1}
  &=\frac{\sin^2 \frac{\pi}{p_2}-2\cos^2\frac{\pi}{\lambda_{D-1}}}{%
    \left(D-3\right)\sin^2\frac{\pi}{ p_2}-2\left(D-4\right)
    \cos^2\frac{\pi}{\lambda_{D-1}}}.
\end{align}
The Regge equations (\ref{eq:chc_ptu}) and 
(\ref{eq:cev_ptu}) give descriptions of the time-development of 
the universe with a regular polytopal Cauchy surface for the parameter 
set $\left[D,\kappa_D,\lambda_D,\mu_D,\zeta_D\right]$.

\begin{figure}[t]
  \centering
  \includegraphics[scale=1]{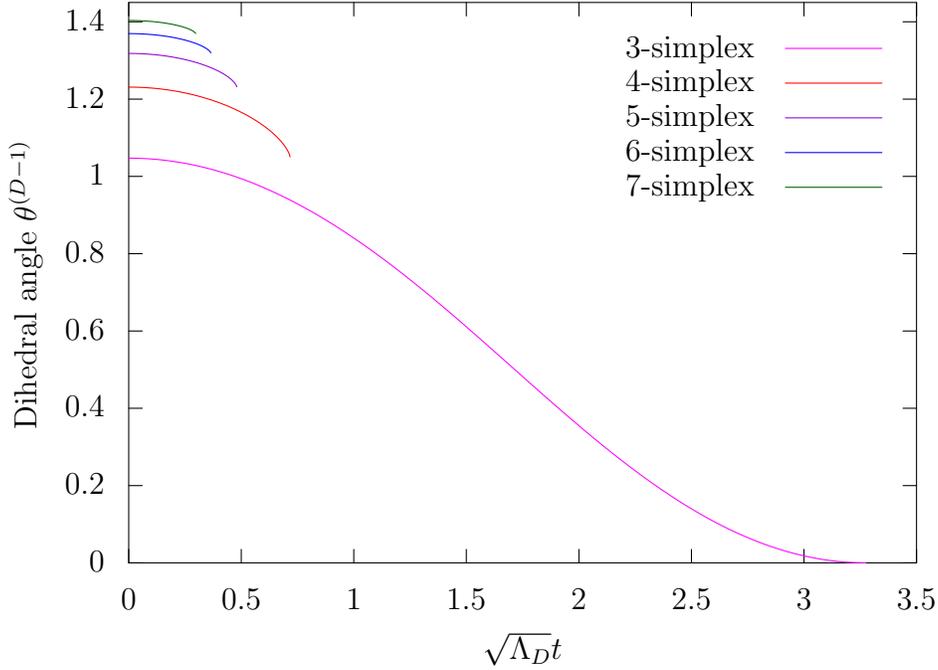}
  \caption{Plots of the dihedral angles of the simplicial polytope 
    models for $3\leq D\leq7$.}
  \label{fig:da_simp}
\end{figure}

Time-development of the dihedral angle $\theta$ can be obtained 
by integrating (\ref{eq:tde_ptu}) numerically for the initial 
condition (\ref{eq:initc}). We give plots of the dihedral angles 
of simplicial polytope models for $D=3,4,\cdots,7$ and 
$0\leq t\leq \tau_\mathrm{p}/2$ in Figure \ref{fig:da_simp}.

\begin{figure}[t]
  \centering
  \includegraphics[scale=1]{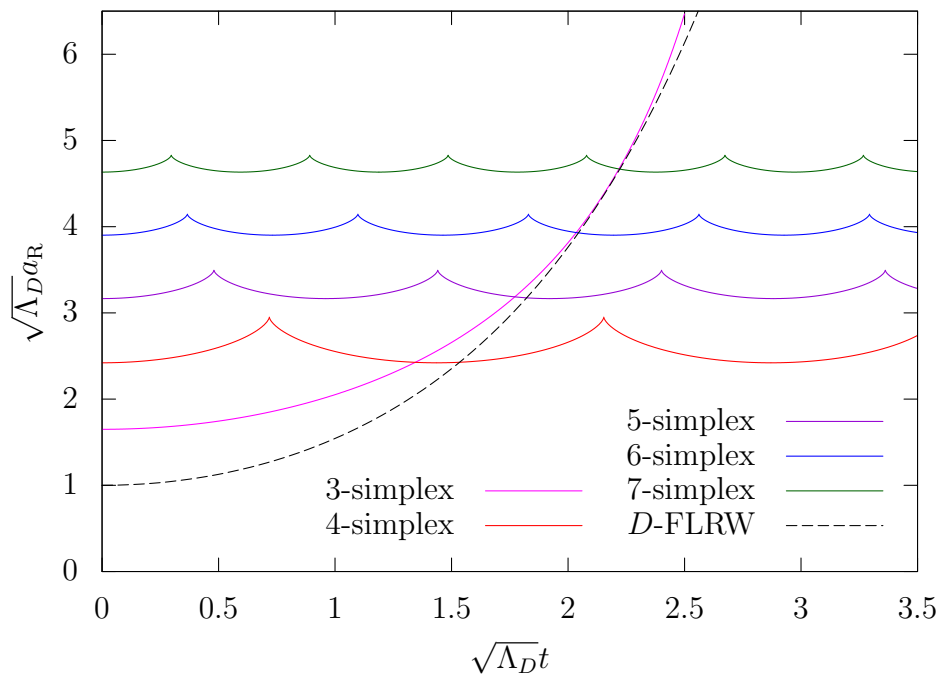}
  \caption{Plots of the scale factors of the simplicial polytope 
    models for $3\leq D\leq7$.
  The broken curve corresponds to the $D$-dimensional FLRW universe.
    }
  \label{fig:sf_simp}
\end{figure}

To compare the polytopal universe with the continuum, we must 
introduce a Regge calculus analog of the scale factor. 
There are, however, ambiguities in defining a radius of a
regular polytope. Here we simply introduce it as the radius of 
the circumsphere of the regular polytope
\begin{align} 
  a_{\mathrm{R}}(t)
      &=\hat{R}_Dl(t).
      \label{eq:sf_ptu}
\end{align}

Inserting the solutions of (\ref{eq:tde_ptu}) into (\ref{eq:sf_ptu}), 
we obtain the time-developments of the scale factors of polytopal 
universes. Figure \ref{fig:sf_simp} shows the behaviors of the simplicial 
universes. The broken curve corresponds to the $D$-dimensional FLRW 
solution. The 3-simplicial model expands faster than the continuum 
one and diverges at $t=\tau_\mathrm{p} / 2$. For $D \geq 4$, after 
arriving at the maximum scale $a(\tau_\mathrm{p}/2)$ the universe begins to 
contract to the initial minimum size $a(0)=a(\tau_\mathrm{p})$. Then the 
universe repeats expanding and contracting with a period $\tau_\mathrm{p}$. 
One easily sees that the $D$-simplices are too crude to approximate the 
continuum solution. The larger the space-time dimensions, the bigger 
difference we have. The situation is somewhat improved by considering 
$D$-orthoplices or $D$-cubes in this order. For fixed space-time 
dimensions the deviation from the continuum FLRW universe become 
smaller as the number of vertices increases. 

In closing this section we comment on the case of $D$-polytopal universe 
without cosmological constant. In this case the Hamiltonian constraint 
(\ref{eq:chc_ptu}) yields $\theta^{(D-1)}=\dfrac{2\pi}{p_D}$. We obtain 
from  (\ref{eq:lstheta}) 
\begin{align}
  \label{eq:hcwc_ptu} 
  \dot{l}^2=-\frac{\cos^2\frac{\vartheta_{D-1}}{2}-\cos^2\frac{\pi}{p_D}}{%
    \hat R_{D-1}^2\cos^2\frac{\vartheta_{D-1}}{2}-\hat R_{D-2}^2
    \cos^2\frac{\pi}{p_D}}
  =-\frac{1}{\hat R_D^2}.
\end{align}
There is no convex regular polytope satisfying this. The Hamiltonian 
constraint, however, admits infinite honeycomb lattices in flat Euclidean 
space. For any space-filling honeycomb the circumradius $\hat R_D$ 
diverges and the dihedral angle is given by 
$\vartheta_D=\pi$, which immediately yields 
\begin{align}
  \label{eq:condsfhl}
  \cos\frac{\vartheta_{D-1}}{2}=\cos\frac{\pi}{p_D}. 
\end{align}
See (\ref{eq:pnvth}). In Table \ref{tab:ssfm} we summarize 
space-filling honeycomb lattices in arbitrary dimensions. 
It is straightforward to verify (\ref{eq:condsfhl}).
We thus obtain static solutions 
$l=\mathrm{const}$. They correspond to the Minkowski 
space-time. In addition, in the case of $\dot{l}^2 > 0$, 
Schl\"afli symbol 
satisfying this inequality stands for a regular lattice of open 
Cauchy surface of constant negative curvature. These results are 
consistent with solutions of the Friedmann equations (\ref{eq:Feq}).
See Table \ref{tab:flrw}.

\begin{table}[t]
\centering
  \begin{tabular}{llll}\hline 
    Dimensions $D$ & Name & Extended Schl\"afli symbol & $[D,\kappa_D,\lambda_D,\mu_D,\zeta_D]$  \\ \hline
    2 & Apeirogon & $\left\{2,\infty , 2 \right\}$ & $[2,\infty,3,3,3]$ \\
    \\
    & Triangular tiling & $\left\{2,3,6,2\right\}$ & $[3,3,6,3,3]$ \\ 
    3 & Square tiling & $\left\{2,4,4,2\right\}$ & $[3,4,4,3,3]$ \\ 
    & Hexagonal tiling & $\left\{2,6,3,2\right\}$ & $[3,6,3,3,3]$ \\ 
    \\
    4 & Cubic honeycomb & $\left\{2,4,3,4,2\right\}$ & $[4,4,3,4,3]$ \\ 
    \\
    & 8-cell honeycomb & $\left\{2,4,3,3,4,2\right\}$ & $[5,4,3,3,4]$ \\
    5 & 16-cell honeycomb & $\left\{2,3,3,4,3,2\right\}$ & $[5,3,3,4,3]$ \\
    & 24-cell honeycomb & $\left\{2,3,4,3,3,2\right\}$ & $[5,3,4,3,3]$ \\ 
    \\
    $n+1\geq6$ & $n$-cubic honeycomb $ \delta_{n+1} $ & $\left\{2,4,3^{n-2},4,2\right\}$
    & $[n+1,4,3,3,4]$ \\ \hline
  \end{tabular}
\caption{Space-filling lattices in Euclidean $\left(D-1\right)$-space.
    The lattices for $ D\geq3 $ are  
    corresponding to Minkowski space-time. 
    The $n$-cubic honeycomb is named by Coxeter as $ \delta_{n+1} $ \cite{Coxeter},
    which has the extended Schl\"afli symbol $ \left\{ 2,4,3^{n-2},4,2 \right\} $.
    The only misfit is $ \delta_2 = \left\{ 2,\infty,2 \right\} $.
    }
  \label{tab:ssfm} 
\end{table}




\section{Fractional Schl\"afli symbol and pseudo-regular \\ $D$-polytopal universes}

\label{sec:prpt} 
\setcounter{equation}{0}

So far we have investigated evolution of regular polytopes 
as a discretized FLRW universe. To go beyond the approximation 
by regular polytopes, we must introduce polytopes with more cells.  
One way to implement this is to employ geodesic domes \cite{TF:2016aa}. 
Hypercube is the only type of regular polytope having subdivisions 
of facets in arbitrary dimensions by the same type of polytopes with
the parent facets. In this section we consider hypercube-based 
geodesic domes as Cauchy surfaces of the universe.

\begin{figure}[t]
  \centering
\includegraphics[scale=1.0]{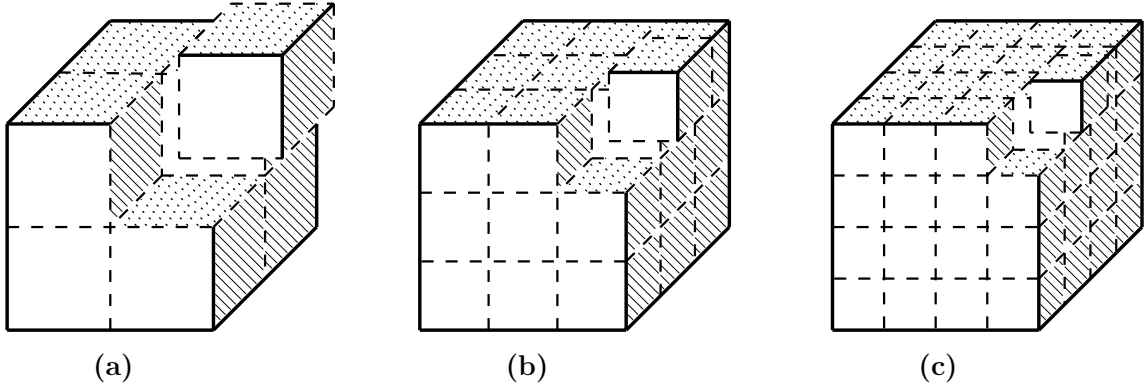}
  \caption{Subdivision of a 3-cube as a cell of a 4-cube for 
    (a) $\nu=2$, (b) $\nu=3$, and (c) $\nu=4$.
    In four dimensions the peaks are the edges.
    Solid lines are the three-way connectors and broken lines 
    the four-way connectors.
    }
  \label{fig:cube_decomp}
\end{figure}

A hypercube in $D$ dimensions has $(D-1)$-cubes as its facets. 
To define a geodesic dome for the hypercube we first divide 
each facet into $\nu^{D-1}$ pieces of $(D-1)$-cubes of edge 
length $l/\nu$ as depicted in Figure \ref{fig:cube_decomp}, 
where $\nu$ is the level of the division, called frequency. 
We then radially project the tessellated hypercube on the 
circumsphere of the original hypercube. This results in a 
tessellation of the circumsphere. The geodesic dome $\Gamma_\nu$ 
can be obtained by replacing each circular arc of the tessellated 
circumsphere with a line segment jointing its end points. 
In general each facet of $\Gamma_\nu$ thus constructed 
is not a flat $(D-1)$-space. We can always decompose these 
facets into flat $(D-1)$-polytopes by adding extra edges. 
The deviations from the flat $(D-1)$-spaces, however, become
negligible as $\nu$ increases. We can effectively regard the 
facets of $\Gamma_\nu$ as flat $(D-1)$-cubes and see any 
polytopal data of the geodesic dome such as the numbers of 
facets, ridges, etc. from the tessellated $D$-cube. 

We can apply Regge calculus 
to $\Gamma_\nu$ as the polyhedral model in Ref. \cite{TF:2016aa}. 
In the infinite frequency limit $ \nu \to \infty $, geodesic 
dome reproduces a smooth sphere. So the model universe approaches 
the FLRW universe in the limit $\nu\rightarrow\infty$.
In practice the larger the frequency, the more cumbersome the Regge 
calculus for geodesic domes becomes. We avoid this complexity by 
introducing pseudo-regular polytopes as in Refs. 
\cite{TF:2016aa,TF:2020aa}.

Let us denote the pseudo-regular polytope corresponding to $\Gamma_\nu$ 
by $\tilde\Gamma_\nu$. We assign it a fractional Schl\"afli symbol 
\begin{align}
  \label{eq:SSprP}
  \{2,4,3^{D-3},p(\nu),2\},
\end{align}
where 
$p(\nu)$ is the averaged number of facets sharing a peak of $\Gamma_\nu$
and the other $D$ integers are the Schl\"afli symbol of the facets of $\Gamma_\nu$.
There are two types of peaks of $\Gamma_\nu$ as illustrated in 
Figure \ref{fig:cube_decomp} for a cell of 4-cube. One is 
shared by three facets. These come from the peaks of the 
original $D$-cube. The other connects four facets.
They are generated in subdividing the facets of the original 
$D$-cube. We refer to the former type as ``three-way connector''
and the later ``four-way connector''. Counting the numbers of 
each type of connectors and averaging the number 
of facets around a peak in $\Gamma_\nu$, we find
\begin{align}
  \label{eq:pnu}
  p(\nu)=\frac{12\nu^2}{3\nu^2+1}.
\end{align}
See Appendix \ref{sec:pnu} for details. The result is independent 
of $D$. Furthermore, the fractional Schl\"afli symbol approaches the 
one of $(D-1)$-cubic honeycomb in the limit $\nu\rightarrow\infty$.

The basic approach of pseudo-regular polytope is to regard 
$\tilde \Gamma_\nu$ as a regular polytope of edge length $l$ 
with the fractional Schl\"afli symbol (\ref{eq:SSprP}) and 
to assume that the model universe is described by the Regge 
equations (\ref{eq:chc_ptu}) and (\ref{eq:cev_ptu}).  
The symbol (\ref{eq:SSprP}) corresponds to the assignment 
\begin{align}
  p_2=4, \quad \lambda_{D-1}=3, \quad p_D=p(\nu).
\end{align}
In particular the normalized circumradii (\ref{eq:RDk}) and 
the dihedral angle (\ref{eq:cosvthD}) coincide with those of 
the regular $D$-cube. They are independent of the frequency $\nu$. 
The differential equation for the dihedral angle $\theta(t)$ can be 
written explicitly as
\begin{align}
  \label{eq:tde_prpt} 
  \dot{\theta}(t)&=\mp\frac{2}{%
    2\pi-p(\nu)\left( \theta(t)-\sin\theta(t)\right)}
  \sqrt{\frac{p(\nu)\Lambda_D\left( 2\pi-p(\nu)\theta(t)\right) \sin2\theta(t)}
    {1-\left(D-2\right)\cos\theta(t)}}\sin\frac{\theta(t)}{2}.
\end{align}
Note that the initial dihedral angle is $\theta(0)=\theta_0
=\vartheta_{D-1}=\pi/2$. Both $\theta_0$ and $\theta_{\mathrm{c}}
=\arccos\dfrac{1}{D-2}$ do not depend on $\nu$. 

The scale factor $a_{\mathrm{R}}$ for the pseudo-regular 
$D$-polytopal universe can be defined similarly as the regular 
polytopal models as 
\begin{align}
  a_{\rm R}\left(t\right)
  \label{eq:sf_prpt_theta}
  &=\hat R_D(\nu)l(t),
\end{align}
where the edge length $l(t)$ for $\tilde\Gamma_\nu$ can be 
found from (\ref{eq:lsqr}) as 
\begin{align}
  \label{eq:elprp}
  l(t)=\frac{2}{\sqrt{\Lambda_D}}
  \sqrt{ \left(\frac{2\pi}{p(\nu)}-\theta(t)\right)\cot\frac{\theta(t)}{2}}.
\end{align}
The normalized circumradius $\hat R_D(\nu)$ also depends on $p_D=p(\nu)$ 
and can be obtained from (\ref{eq:hRD}) as  
\begin{align}
\label{eq:Rhat_dcube}
  \hat R_D(\nu)=\frac{1}{2}
  \sqrt{D-2-\sec\frac{2\pi}{p(\nu)}}.
\end{align}
For $\nu=1$ this coincides with the circumradius of a regular $D$-cube 
of unit edge length. It grows with the frequency $\nu$ and diverges linearly
for $\nu\rightarrow\infty$. 
In fact Eq. (\ref{eq:Rhat_dcube}) can be approximated for large frequency by
\begin{align}
\hat{R}_D (\nu) \approx \sqrt{ \frac{3}{2 \pi} } \nu.
\end{align}
On the other hand the edge length 
(\ref{eq:elprp}) decreases roughly inversely with $\nu$ and approaches 
zero as $\nu\rightarrow\infty$. This can be seen explicitly for the 
initial edge length
\begin{align}
  \label{eq:initelprp}
  l(0)=\sqrt{\frac{2\pi}{3\Lambda_D}}\frac{1}{\nu}.
\end{align}
The scale factor (\ref{eq:sf_prpt_theta}), however, remains finite for 
$\nu\rightarrow\infty$. 
Noting that $\hat R_{D-k}$ ($k=1,2,3$) 
are independent of $\nu$ as given by (\ref{eq:RDk}),
it is 
straightforward to verify that the Regge equations (\ref{eq:chc_ptu}) 
and (\ref{eq:cev_ptu}) for $\tilde\Gamma_\nu$ reduce to the Friedmann 
equations (\ref{eq:Feq}) in the limit $\nu\rightarrow\infty$. 

\begin{figure}[t]
  \centering
  \includegraphics[scale=1]{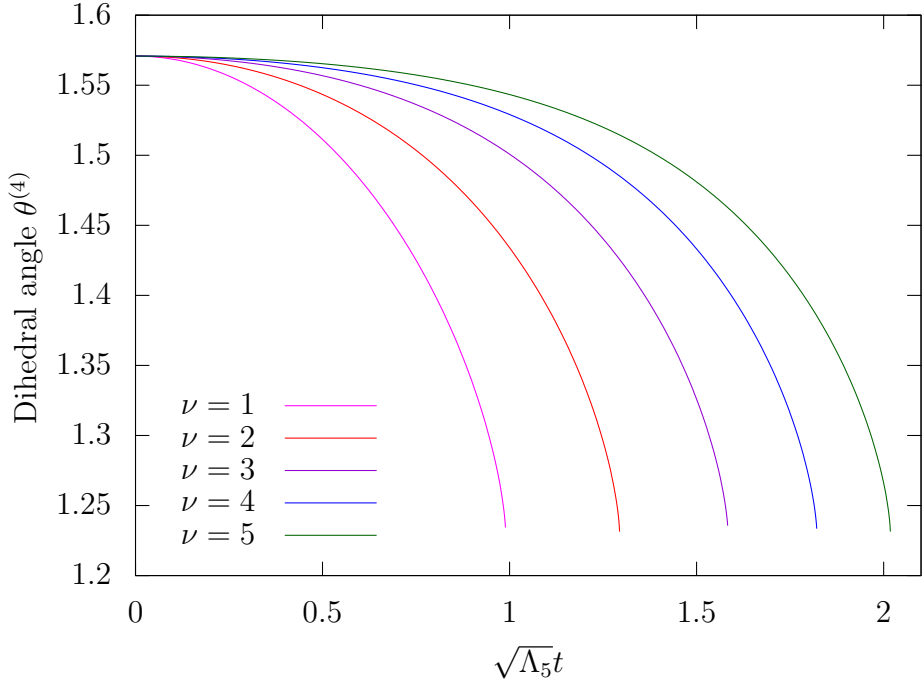}
  \caption{Plots of the dihedral angles of the pseudo-regular 
    5-polytopal universes for $\nu\leq5$.}
  \label{fig:da_prpt}
\end{figure}

\begin{figure}[t]
  \centering
\includegraphics[scale=1]{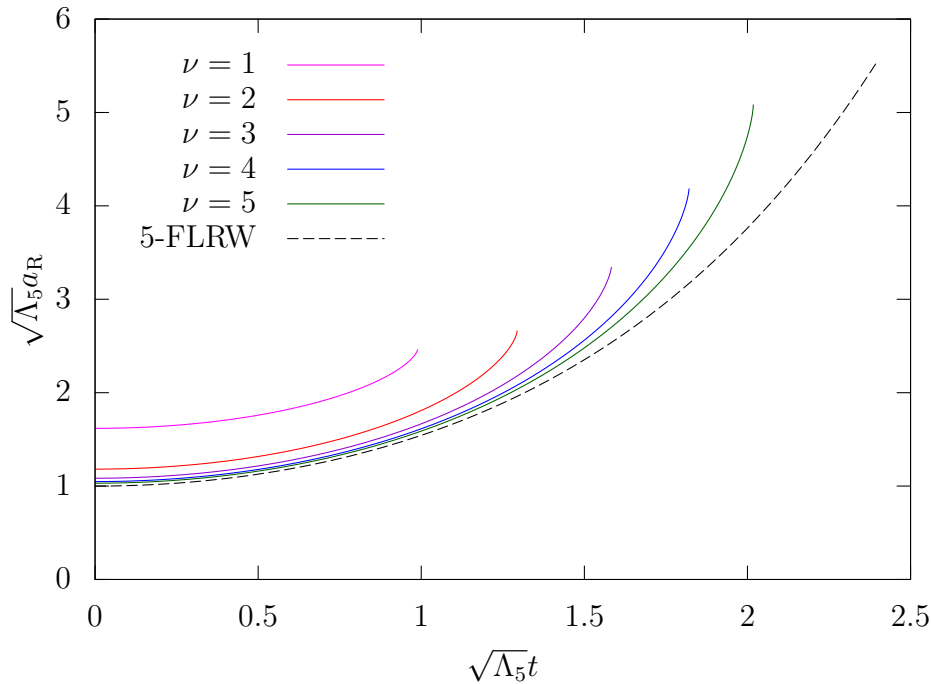}
  \caption{Plots of the scale factors of the pseudo-regular 5-polytopal universes for $\nu \leq 5$.
  The broken curve corresponds to the five-dimensional FLRW universe.
    }
  \label{fig:sf_prpt}
\end{figure}

\begin{figure}[t]
  \centering
\includegraphics[scale=1]{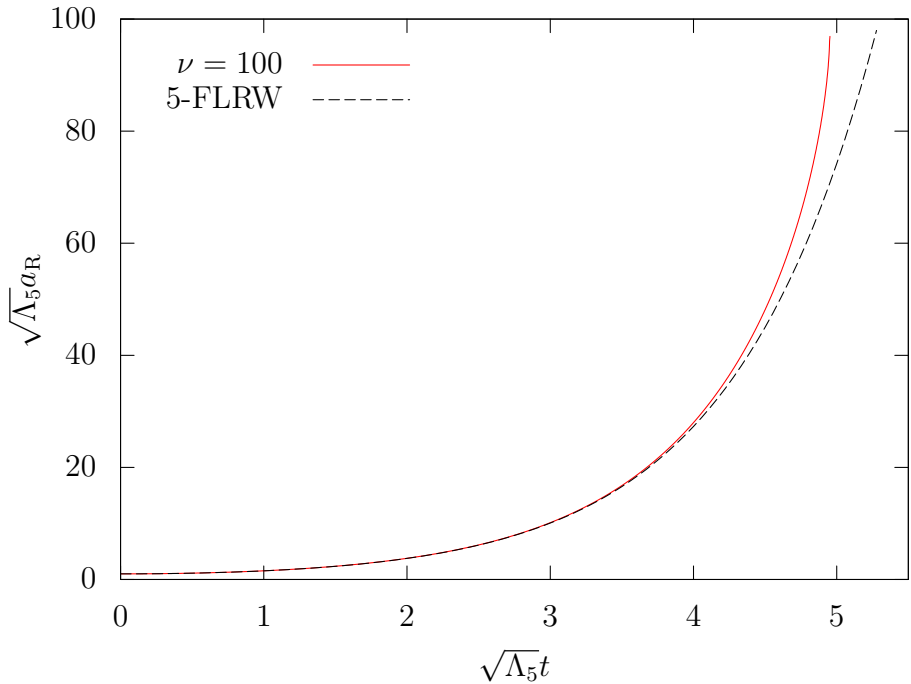}
  \caption{Plot of the scale factor of the pseudo-regular 5-polytopal universe for $\nu=100$.
  The broken curve stands for the exact solution of the continuum theory.
    }
  \label{fig:sf_prpt_nu100}
\end{figure}

To see the dependences on $\nu$ we give plots of the dihedral angles
in Figure \ref{fig:da_prpt} and those of the scale factors in 
Figure \ref{fig:sf_prpt} for $D=5$, $\nu\leq 5$, and 
$0\leq t\leq\tau_\mathrm{p}(\nu)/2$, where $\tau_\mathrm{p}(\nu)$ is 
the period of the 
oscillation of $\tilde\Gamma_\nu$. One might think that $D$-cube-based 
pseudo-regular polytopes are too crude to approximate $D$-spheres. 
As can be seen from Figure \ref{fig:sf_prpt}, the scale factor 
approaches rapidly the continuum one as $\nu$ increases. 
As mentioned above, the geodesic dome $\Gamma_\nu$ becomes impractical to 
carry out Regge calculus for large $\nu$. The advantage of the approach 
of pseudo-regular polytopes is its applicability to arbitrarily large 
frequency without effort. The scale factor for $\nu = 100$ is shown in 
Figure \ref{fig:sf_prpt_nu100}. Coincidence with the continuum theory 
is excellent for $\sqrt{\Lambda_5}t\sim4$. The edge length becomes 
comparable with $1/\sqrt{\Lambda_5}$ at around $\sqrt{\Lambda_5}t\sim4$,   
onset of the deviation from the continuum solution.

\section{Summary and discussions}

\label{sec:sum}
\setcounter{equation}{0}

Following the CW formalism, we have carried out Regge calculus 
for closed FLRW universe with a positive cosmological constant 
in arbitrary dimensions.  The geometrical characterization of 
regular polytopes by the Schl\"afli symbol has turned out to be 
very efficient in describing systematically the discrete FLRW 
universe in spite of there being only three types of regular 
polytopes in dimensions more than four. We have given the Regge 
action in closed form in the continuum time limit. It possesses 
a reparameterization invariance of time variable to ensure 
coordinate independence of the formalism. The Regge equations 
are the Hamiltonian constraint and the evolution equation as 
the continuum theory, describing the time development of the 
discrete FLRW universe. They coincide with the previous results 
in three and four dimensions \cite{TF:2016aa,TF:2020aa}. In 
particular under the gauge choice (\ref{eq:gconn}) the 
circumsphere of the regular polytope repeats periodically 
expansion and shrinking in any dimensions larger than four
as the four dimensional case. The Regge equations have more or 
less the same structures in dimensions greater than three. 
It is only in three dimensions where the edge length diverges 
in finite time. 

As we have shown in Sect. \ref{sec:req} the approximation by 
regular polytopes is not so accurate even for $\sqrt{\Lambda_D}t\ll 1$. 
The situation gets worse as the dimensions increase. 
This is contrasted with the cases of dodecahedron in three 
dimensions and 120-cell in four dimensions, which describe 
the continuum FLRW universe rather well until $t$ becomes 
comparable with $1/\sqrt{\Lambda_D}$. The difference basically 
comes from that of the number of vertices in a polytope. A 
120-cell has six hundred vertices, whereas a 4-cube does only sixteen. 
In five or more dimensions there are no such special 
polytopes. 
One must refine the tessellation of the Cauchy 
surface by 
nonregular polytopes with smaller cells
to have better 
approximations. Though this can be done by extending the 
geodesic domes in three dimensions, we have analyzed 
pseudo-regular polytopes with the expectation that the Regge 
equations for the pseudo-regular polytopes approximate well 
the Regge calculus of the corresponding geodesic domes. 
We stress that the pseudo-regular polytope is a substitute 
of the corresponding geodesic domes characterized by 
the frequency $\nu$, not the continuum hypersphere. 
The Regge equations (\ref{eq:chc_ptu}) and (\ref{eq:cev_ptu}) 
therefore should be considered as an effective description 
of the Regge equations for the geodesic dome, not of the 
continuum Freedman equations. The approach of pseudo-regular 
polytopes can be applied to an arbitrary $\nu$. 
In particular 
we can infer the validity of Regge calculus for geodesic 
domes. Because of this, the pseudo-regular polytope universe 
begins to deviate from the continuum solution when the edge 
length becomes larger than $1/\sqrt{\Lambda_D}$. 

In this paper we have considered vacuum universes without 
matters. Incorporating gravitating matter sources is worth
investigation. In General Relativity, Friedmann equations 
have a solution for a negative cosmological constant. It 
describes hyperbolic Cauchy surfaces expanding or contracting 
with time. Applying the method of pseudo-regular polytope to 
such non-compact universe be interesting. We will address 
these issues elsewhere.

\vskip .5cm




\appendix




\section{Circumradii and dihedral angles of 
regular polytopes}
\label{sec:cada}
\setcounter{equation}{0}

In this appendix we give closed expressions for circumradius 
and dihedral angle of an arbitrary regular polytope $\Pi_n
= \left\{ p_1 , \cdots , p_n , p_0 \right\} $ in $n$ 
dimensions \cite{Coxeter,Hitotsumatsu}. By definition $\Pi_n$ has 
regular $(n-1)$-polytopes $\Pi_{n-1}
= \{p_1,\cdots,p_{n-1} , p_0\} $ as its $N^{(n)}_{n-1}$ cells.
Each cell of $\Pi_n$ also has $N^{(n-1)}_{n-2}$ 
cells.  
They are regular $(n-2)$-polytopes $\Pi_{n-2}
= \{p_1,\cdots,p_{n-2},p_0\} $. Similar decomposition of 
a daughter cell of a parent cell can be continued until 
we arrive at the vertices of the original $n$-polytope. They are 
zero-dimensional cells $\Pi_0=\{p_0\}$. Incidentally $\Pi_1
= \{p_1,p_0\} $ corresponds to edges. 

\begin{figure}[t]
  \centering
  \includegraphics[scale=.9]{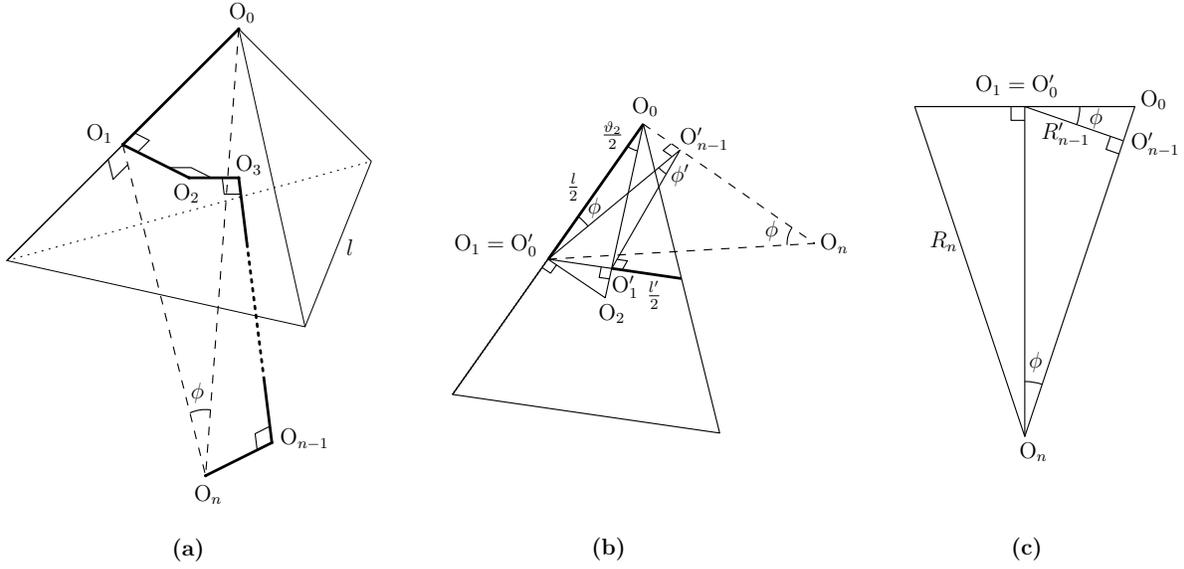}
  \caption{
  (a)
  Centers of circumspheres of the $\Pi_k ~ \left( k = 0 , 1 , \cdots , n \right)$
  in the case of $\Pi_3 = \left\{ 2,3,3,2 \right\} $,
  where $n$ is the dimension of original regular polytope $\Pi_n$.
  Obviously $\mathrm{O}_1$ is just the midpoint of $\Pi_1$ and $\mathrm{O}_0 = \Pi_0$.
  (b)
  $\mathrm{O}'_0 , \mathrm{O}'_1 , \cdots , \mathrm{O}'_{n-1}$ are the centers in the vertex figure $\Pi'_{n-1}$
  and 
  (c)
  $R_n$ and $R'_{n-1}$ are the circumradii of $\Pi_n$ and $\Pi'_{n-1}$, respectively.
  }
  \label{fig:centers}
\end{figure}

We now choose a set of cells $\Pi_0$, $\Pi_1$, $\cdots$, 
$\Pi_{n-1}$ satisfying $\Pi_0\subset\Pi_1\subset\cdots
\subset\Pi_{n-1}\subset \Pi_n$ and denote the centers of 
circumspheres of the $\Pi_k$ by $\mathrm{O}_k$ ($k=0,1,\cdots,n$). 
See Figure \ref{fig:centers}(a). 
Then $R_n=\overline{\mathrm{O}_n\mathrm{O}_0}$ is the 
circumradius of $\Pi_n$. It is given by
\begin{align}
  \label{eq:hatRn}
  R_n=l \hat R_n \quad \hbox{with} \quad \hat R_n=\frac{1}{2}\csc\phi,
\end{align}
where 
$l$ is the edge length of the original regular polytope $\Pi_n$ and
the angle $\phi$ is defined by
\begin{align}
\phi=\angle\mathrm{O}_0\mathrm{O}_n\mathrm{O}_1.
\end{align}
Note that the line 
segments connecting $\mathrm{O}_k$ and $\mathrm{O}_{k+1}$ 
($k=0,1,\cdots,n-1$) are orthogonal to one another.

We next consider the section obtained by cutting 
out $\Pi_n$ by the hyperplane through the centers 
of edges meeting at $\mathrm{O}_0$. 
It is a regular $(n-1)$-polytope $\Pi'_{n-1}= \{p_1,p_3,\cdots,p_n,p_0\} $, 
called vertex figure,
of an edge length $l'=l\cos\dfrac{\pi}{p_2}$. We can 
pick up a sequence of centers $\mathrm{O}_0'(=\mathrm{O}_1)$, 
$\mathrm{O}_1'$, $\mathrm{O}_2'$, $\cdots$, $\mathrm{O}_{n-1}'$ 
in $\Pi_{n-1}'$ as we have done for $\Pi_n$. 
See Figure \ref{fig:centers}(b).
The circumradius 
of $\Pi'_{n-1}$ is given by $R_{n-1}'=\dfrac{l'}{2}\csc\phi'
=\dfrac{l}{2}\csc\phi'\cos\dfrac{\pi}{p_2}$ with 
$\phi'=\angle\mathrm{O}_0'\mathrm{O}_{n-1}'\mathrm{O}_1'$. 
We can also write it as $R_{n-1}'=\dfrac{l}{2}\cos\phi$ since  
$\triangle\mathrm{O}_0\mathrm{O}_1\mathrm{O}_{n-1}'$ is similar to 
the right triangle $\triangle\mathrm{O}_0\mathrm{O}_1\mathrm{O}_n$
as depicted in Figure \ref{fig:centers}(c).
These lead to a constraint between the angles $\phi$ of 
$\Pi_n= \{p_1,p_2,\cdots,p_n,p_0\} $ and $\phi'$ of 
$\Pi'_{n-1}= \{p_1,p_3,\cdots,p_n,p_0\} $ as
\begin{align}
  \label{eq:sinphi}
  \cos\phi=\csc\phi'\cos\frac{\pi}{p_2} \quad\hbox{or}\quad
  \sin^2\phi=1-\frac{\cos^2\frac{\pi}{p_2}}{\sin^2\phi'}. 
\end{align}
We thus obtain $\sin^2\phi$ in terms of a continued fraction as 
\begin{align}
  \label{eq:cfsinphi}
  \sin^2\phi=1-\frac{\cos^2\frac{\pi}{p_2}}{1-}
  \frac{\cos^2\frac{\pi}{p_3}}{1-}\cdots
  \frac{\cos^2\frac{\pi}{p_{n-2}}}{1-}
  \frac{\cos^2\frac{\pi}{p_{n-1}}}{\sin^2\frac{\pi}{p_n}}. 
\end{align}

More tractable expressions for the normalized circumradius 
$\hat R_n$ can be found by applying the substitution rules\cite{TF:2020aa}
\begin{align}
\label{eq:rrcirc} 
\begin{cases}
  \sin^2\frac{\pi}{p_{k-1}}&\to~\sin^2\frac{\pi}{p_{k-1}}
  \sin^2\frac{\pi}{p_{k+1}} \\
  \cos^2\frac{\pi}{p_{k-1}}&\to~\cos^2\frac{\pi}{p_{k-1}}
  \sin^2\frac{\pi}{p_{k+1}} \\
  \sin^2\frac{\pi}{p_{k-2}}&\to~\sin^2\frac{\pi}{p_{k-2}}
  \left(1-\csc^2\frac{\pi}{p_k}\cos^2\frac{\pi}{p_{k+1}}\right) \\
  \cos^2\frac{\pi}{p_{k-2}}&\to~\cos^2\frac{\pi}{p_{k-2}}
  \left(1-\csc^2\frac{\pi}{p_k}\cos^2\frac{\pi}{p_{k+1}}\right)
\end{cases}
\end{align}
to $\hat{R}_k$
with the initial condition
\begin{align}
  \label{eq:iccirc} 
  \hat{R}_1= \frac{1}{2}\sqrt{\frac{\sin^2\frac{\pi}{p_1}}{\sin^2\frac{\pi}{p_0}}}.
\end{align}
For reader's reference we give the next four of the circumradii
\begin{align}
  \hat R_2&=\frac{1}{2}\sqrt{\frac{\sin^2\frac{\pi}{p_1}}{%
      \sin^2\frac{\pi}{p_0}\sin^2\frac{\pi}{p_2}}}, \\
  \hat R_3&=\frac{1}{2}\sqrt{\frac{\sin^2\frac{\pi}{p_1}
      \sin^2\frac{\pi}{p_3}}{\sin^2\frac{\pi}{p_0}
      \left(\sin^2\frac{\pi}{p_2}-\cos^2\frac{\pi}{p_3}\right)}},\\
  \hat R_4&=\frac{1}{2}\sqrt{\frac{\sin^2\frac{\pi}{p_1}
      \left(\sin^2\frac{\pi}{p_3}-\cos^2\frac{\pi}{p_4}\right)}{%
      \sin^2\frac{\pi}{p_0}\left(\sin^2\frac{\pi}{p_2}
        \sin^2\frac{\pi}{p_4}-\cos^2\frac{\pi}{p_3}\right)}}, \\
  \hat R_5&=\frac{1}{2}\sqrt{\frac{\sin^2\frac{\pi}{p_1}
      \left(\sin^2\frac{\pi}{p_3}\sin^2\frac{\pi}{p_5}
        -\cos^2\frac{\pi}{p_4}\right)}{%
      \sin^2\frac{\pi}{p_0}\left(\sin^2\frac{\pi}{p_2}
        \left(\sin^2\frac{\pi}{p_4}-\cos^2\frac{\pi}{p_5}\right)
        -\cos^2\frac{\pi}{p_3}\sin^2\frac{\pi}{p_5}\right)}}, 
\end{align} 
where $p_0=p_1=2$. 
One can easily see that (\ref{eq:cfsinphi}) reproduces the same 
results.

\begin{figure}[t]
  \centering
  \includegraphics[scale=1]{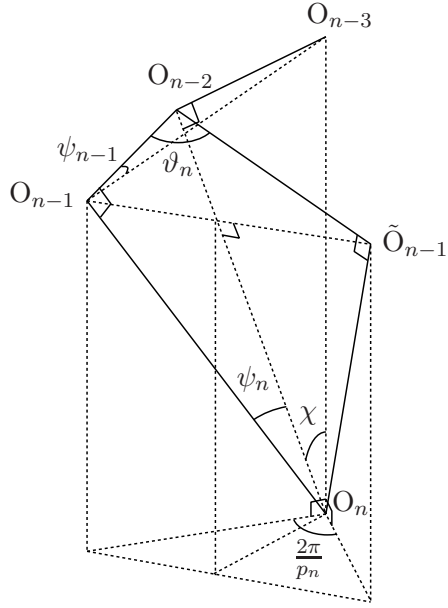}
  \caption{$\mathrm{O}_n$ is the circumcenter of $\Pi_n$.
  $\mathrm{O_{n-1}}$ and $\mathrm{\tilde{O}}_{n-1}$ are the centers of two facets sharing a ridge $ \Pi_{n-2} $ centered at $ \mathrm{O}_{n-2} $,
  and
  $\mathrm{O}_{n-3}$ is located at the center of a peak $\Pi_{n-3}$ included in the $\Pi_{n-2}$.
  }
  \label{fig:DHA}
\end{figure}

Returning to the set of points $\mathrm{O}_0$, $\mathrm{O}_1$, 
$\cdots$, $\mathrm{O}_n$ in $\Pi_n$, we define angles 
\begin{align}
  \label{eq:psichi}
  \psi_n=\angle\mathrm{O}_{n-2}\mathrm{O}_n\mathrm{O}_{n-1}, \quad
  \psi_{n-1}=\angle\mathrm{O}_{n-3}\mathrm{O}_{n-1}\mathrm{O}_{n-2}, \quad
  \chi=\angle\mathrm{O}_{n-3}\mathrm{O}_n\mathrm{O}_{n-2}. 
\end{align}
See Figure \ref{fig:DHA}. 
The dihedral angle $\vartheta_n$ between 
the two facets of $\Pi_n$ connected at the ridge $\Pi_{n-2}$
is related to $\psi_n$ by
\begin{align}
  \label{eq:vthtopsi}
  \vartheta_n=\pi-2\psi_n. 
\end{align}
To see this consider a pair of adjacent 
facets of $\Pi_n$, one is the 
$\Pi_{n-1}$ centered at $\mathrm{O}_{n-1}$ and the other centered 
at  $\tilde{\mathrm{O}}_{n-1}$. The four points $\mathrm{O}_n$, 
$\mathrm{O}_{n-1}$, $\mathrm{O}_{n-2}$, $\tilde{\mathrm{O}}_{n-1}$ with 
$\angle\mathrm{O}_{n-2}\mathrm{O}_{n-1}\mathrm{O}_n
=\angle\mathrm{O}_{n-2}\tilde{\mathrm{O}}_{n-1}\mathrm{O}_n=\dfrac{\pi}{2}$ 
lie on a two dimensional plane, from which (\ref{eq:vthtopsi}) immediately 
follows.

There are $p_n$ such planes around the axis $\mathrm{O}_{n-3}\mathrm{O}_n$. 
This implies that the projections of $\mathrm{O}_{n-1}\mathrm{O}_n$ and 
$\tilde{\mathrm{O}}_{n-1}\mathrm{O}_n$ onto the plane perpendicular 
to  $\mathrm{O}_{n-3}\mathrm{O}_n$ in the three dimensional space 
containing $\mathrm{O}_{n-3}$, $\mathrm{O}_{n-2}$, $\mathrm{O}_{n-1}$, 
and $\mathrm{O}_n$ make an angle $\dfrac{2\pi}{p_n}$. It can be seen 
that the three angles (\ref{eq:psichi}) satisfy 
$\tan\chi=\sin\psi_n\tan\psi_{n-1}$ and $\tan\psi_n
=\sin\chi\tan\dfrac{\pi}{p_n}$, from which we obtain 
\begin{align}
  \label{eq:ptodha}
  \sin\psi_{n-1}\cos\psi_n=\cos\frac{\pi}{p_n}. 
\end{align}
This enables us to express $p_n$ in terms of the dihedral angles 
as
\begin{align}
\label{eq:pnvth}
  p_n=\frac{\pi}{\arccos\left(\cos\frac{\vartheta_{n-1}}{2}%
    \sin\frac{\vartheta_n}{2}\right)}. 
\end{align}

Eq. (\ref{eq:ptodha}) can be written as 
\begin{align}
  \label{eq:sinpsi}
  \sin^2\psi_n=1-\frac{\cos^2\frac{\pi}{p_n}}{\sin^2\psi_{n-1}}. 
\end{align}
One easily recognize similarity to (\ref{eq:sinphi}). We 
thus arrive at an expression for the dihedral angle $\vartheta_n$ 
in terms of a continued fraction as 
\begin{align}
  \label{eq:cfvthn}
  \vartheta_n=2\arcsin\sqrt{\frac{\cos^2\frac{\pi}{p_n}}{1-}
  \frac{\cos^2\frac{\pi}{p_{n-1}}}{1-}\cdots
  \frac{\cos^2\frac{\pi}{p_4}}{1-}
  \frac{\cos^2\frac{\pi}{p_3}}{\sin^2\frac{\pi}{p_2}}}.
\end{align}
Eq. (\ref{eq:vthtopsi}) for $n=2$ gives 
\begin{align}
\label{eq:vth2}
\vartheta_2=\pi-\frac{2\pi}{p_2} = 2 \arcsin \left( \cos \frac{\pi}{p_2} \right),
\end{align}
which is the interior angle of a regular polygon $\{p_1,p_2,p_0\}$. 
The next three dihedral angles are explicitly given by
\begin{align}
  \vartheta_3&=2\arcsin\frac{\cos\frac{\pi}{p_3}}{\sin\frac{\pi}{p_2}}, \\
  \vartheta_4&=2\arcsin\frac{\sin\frac{\pi}{p_2}\cos\frac{\pi}{p_4}}{%
    \sqrt{\sin^2\frac{\pi}{p_2}-\cos^2\frac{\pi}{p_3}}}, \\
  \vartheta_5&=2\arcsin\left(\sqrt{\frac{\sin^2\frac{\pi}{p_2}
        -\cos^2\frac{\pi}{p_3}}{\sin^2\frac{\pi}{p_2}
        \sin^2\frac{\pi}{p_4}-\cos^2\frac{\pi}{p_3}}}
    \cos\frac{\pi}{p_5}\right).
\end{align}
Moreover Eq. (\ref{eq:pnvth}) with (\ref{eq:vth2}) gives a natural extension of $\vartheta_n$ for a 1-polytope as
\begin{align}
\label{eq:vth1}
\vartheta_1 = 0.
\end{align}

It is possible to write the circumradius in terms of the 
dihedral angles without using continued fraction. To do this let us 
denote the distance between 
$\mathrm{O}_{k-1}$ and $\mathrm{O}_k$ by $d_k$ ($k=1,\cdots,n$), 
then $d_{k+1}=d_k\tan\dfrac{\vartheta_{k+1}}{2}$ with 
$d_1=\dfrac{l}{2}$. The square of normalized circumradius can 
be expressed as 
\begin{align}
  \label{eq:rntodha}
  \hat R_n^2=\frac{1}{l^2}\sum_{j=1}^n d_j^2
  =\frac{1}{4}\left(1+\sum_{j=2}^n\prod_{k=2}^{j}
    \tan^2\frac{\vartheta_k}{2}\right) \qquad (n\geq2).
\end{align}
It is easy to show the following recurrence relations  
\begin{align}
  \label{eq:Rnrecr}
  &\hat R_n^2-\hat R_{n-1}^2=(\hat R_{n-1}^2
  -\hat R_{n-2}^2)\tan^2\frac{\vartheta_n}{2}, \\
  \label{eq:RntoDHA}
  &\frac{\hat R_{n-2}^2}{\hat R^2_{n-1}}\sin^2\frac{\vartheta_n}{2}
  =\frac{\cos^2\frac{\pi}{p_n}}{%
    1-\dfrac{\hat R_{n-3}^2}{\hat R^2_{n-2}}\sin^2\dfrac{\vartheta_{n-1}}{2}}. 
\end{align}
The second of these leads to 
\begin{align}
  \label{eq:maxangl}
  \frac{\hat R_{n-2}^2}{\hat R^2_{n-1}}\sin^2\frac{\vartheta_n}{2}
  =\frac{\cos^2\frac{\pi}{p_n}}{
    1-}\frac{\cos^2\frac{\pi}{p_{n-1}}}{%
    1-}\cdots\frac{\cos^2\frac{\pi}{p_4}}{%
  \sin^2\frac{\pi}{p_3}}. 
\end{align}
Comparing this with (\ref{eq:cfvthn}), we see 
$2\arcsin\left(\dfrac{\hat R_{n-2}}{\hat R_{n-1}}
\sin\frac{\vartheta_n}{2}\right)$ coincides with a 
dihedral angle 
of a regular $(n-1)$-polytope $\{p_1,p_3,\cdots,p_n,p_0\}$. 
It is a vertex figure of $\Pi_n$.

In six or higher dimensions every convex regular polytope and space-filling lattice have Schl\"afli 
symbol $ p_5 = \cdots = p_{n-1} = 3 $ in common as given in 
Tables \ref{tab:ssfrpt} and \ref{tab:ssfm}. Therefore in these dimensions the circumradius 
and the dihedral angle
of a regular $n$-polytope
depend on only five parameters  
$p_2$, $p_3$, $p_4$, $p_n$, and $n$. Inserting $p_0 = p_1 = 2$ and 
$ p_5 = \cdots = p_{n-1} = 3 $ into (\ref{eq:cfsinphi}) and 
(\ref{eq:cfvthn}), we obtain the general forms of the 
circumradius and the dihedral angle of a unit equilateral 
polytope for $n \geq 5$ as
\begin{align}
\label{eq:hatRD}
\hat{R}_n &= \frac{1}{2} \sqrt{ \frac{ \left[ 1 - \left( D-4 \right) \cos \frac{ 2 \pi }{ p_n } \right] \sin^2 \frac{\pi}{p_3} - 2 \left[ 1 - \left( D - 5 \right) \cos \frac{ 2 \pi }{ p_n } \right] \cos^2 \frac{ \pi }{ p_4  } }{%
\left[ 1 - \left( D-4 \right) \cos \frac{2 \pi}{p_n} \right] \left( \sin^2 \frac{\pi}{p_3} - \cos^2 \frac{\pi}{p_2}  \right) 
- 
2 \left[ 1 - \left( D-5 \right) \cos \frac{2 \pi}{p_n} \right] \sin^2 \frac{\pi}{p_2} \cos^2 \frac{ \pi}{p_4} 
} }, \\
\label{eq:vTD}
\vartheta_n &= 2 \arcsin \left( \sqrt{ 2 \frac{ \sin^2 \frac{ \pi }{ p_2 } \left[ 1 - \left( D - 5 \right) \cos \frac{ 2 \pi }{ p_4 } \right] - \left( D - 4 \right) \cos^2 \frac{\pi}{ p_3 } }{%
\sin^2 \frac{ \pi }{ p_2 } \left[ 1 - \left( D - 4 \right) \cos \frac{ 2 \pi }{ p_4 } \right] - \left( D - 3 \right) \cos^2 \frac{\pi}{ p_3 }
} } \cos \frac{\pi}{ p_n } \right).
\end{align}




\section{Dihedral angles of a $(D-1)$-polytopal frustum}
\label{sec:dadpf}
\setcounter{equation}{0}

\begin{figure}[t]
  \centering
  \includegraphics[scale=1]{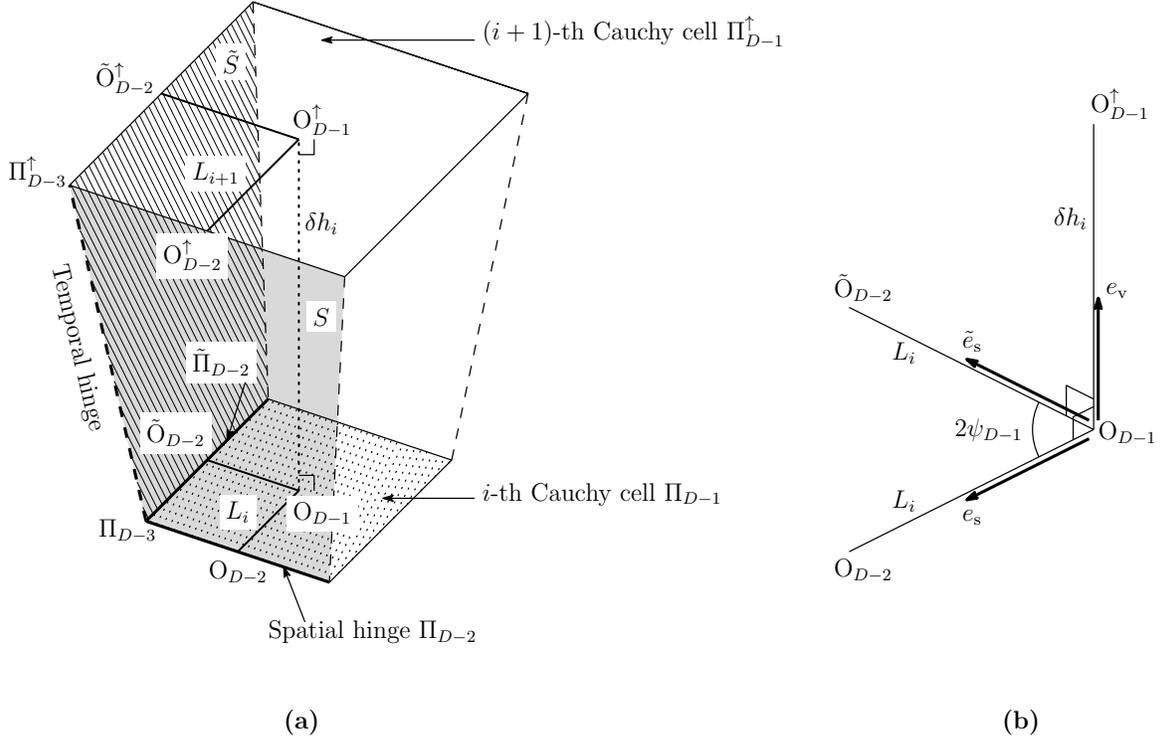}
  \caption{
  (a) $(D-1)$-polytopal frustum 
  and (b) three unit vectors $e_\mathrm{s}$, $\tilde{e}_\mathrm{s}$, and $e_\mathrm{v}$ 
  parallel to $\mathrm{O}_{D-1} \mathrm{O}_{D-2}$, $\mathrm{O}_{D-1} \tilde{\mathrm{O}}_{D-2}$, 
  and $\mathrm{O}_{D-1} \mathrm{O}^\uparrow_{D-1}$, respectively.
  }
  \label{fig:frustum_da}
\end{figure}

In this appendix we give a derivation of the dihedral 
angles (\ref{eq:theta_ptu}) and (\ref{eq:phi_down_ptu}). 

We first consider a temporal hinge. Let us choose an 
$i$-th $(D-1)$-polytopal frustum with $\Pi_{D-1}^\uparrow$ 
and $\Pi_{D-1}$ as 
the upper and lower cells as illustrated schematically 
in Figure \ref{fig:frustum_da}(a). 
The hinge is supposed to contain 
the regular $(D-3)$-polytopes $\Pi_{D-3}$ 
and $\Pi^\uparrow_{D-3}$.
Each vertex of $\Pi_{D-3}$ is connected with the corresponding vertex 
of $ \Pi^\uparrow_{D-3} $ by a strut.
The height of the frustum $\delta h_i$ 
and 
the distance $L_i$ between $\mathrm{O}_{D-1}$ and $\mathrm{O}_{D-2}$ 
are given by  
\begin{align}
  \label{eq:Ldh}
  \delta h_i=\sqrt{m_i^2-\hat R_{D-1}^2\delta l_i^2}, \qquad  
  L_i=\sqrt{\hat R_{D-1}^2-\hat R_{D-2}^2}~l_i. 
\end{align}   

As illustrated in 
Figure \ref{fig:frustum_da}(a) there are two lateral cells 
jointed at the temporal hinge, one containing $\Pi_{D-2}$ 
and the other containing $\tilde\Pi_{D-2}$. Let $S$ be 
the $(D-1)$-dimensional hyperplane containing 
$\Pi_{D-2}$ and $\Pi^\uparrow_{D-3}$, then the 
outgoing unit normal to $S$ can be written as
\begin{align}
  u=\frac{ \delta h_i e_\mathrm{s} - \delta L_ie_\mathrm{v}}{%
    \sqrt{\delta h_i^2+\delta L_i^2}}
\end{align}
with $\delta L_i=\sqrt{\hat R_{D-1}^2-\hat R_{D-2}^2}\delta l_i$, 
where $e_\mathrm{s}$ and $e_\mathrm{v}$ are,
as depicted in Figure \ref{fig:frustum_da}(b),
the unit vectors parallel to 
$\mathrm{O}_{D-1}\mathrm{O}_{D-2}$ and 
$\mathrm{O}_{D-1}\mathrm{O}_{D-1}^\uparrow$, respectively. 
Likewise, we set up the hyperplane $\tilde{S}$ as the section containing 
$\tilde{\Pi}_{D-2}$ and $\Pi^\uparrow_{D-3}$. 
Then the outgoing unit normal to $\tilde S$ takes the form
\begin{align}
  \tilde u=\frac{ \delta h_i \tilde e_\mathrm{s} - \delta L_i e_\mathrm{v}}{%
    \sqrt{\delta h_i^2+\delta L_i^2}},
\end{align}
where $\tilde e_\mathrm{s}$ is the unit vector parallel to 
$\mathrm{O}_{D-1}\tilde{\mathrm{O}}_{D-2}$. 
Since 
$e_\mathrm{s}\cdot \tilde e_\mathrm{s}=\cos2\psi_{D-1}=-\cos\vartheta_{D-1}$, 
we can find the dihedral angle $\theta^{(D-1)}_i$ from  
\begin{align}
  \cos\theta^{(D-1)}_i=-u\cdot \tilde u
  =\frac{\delta h_i^2\cos\vartheta_{D-1}-\delta L_i^2}{%
    \delta h_i^2+\delta L_i^2}.
\end{align}
Eq. (\ref{eq:theta_ptu}) follows from this.

We next turn to the dihedral angle between the 
two cells meeting at the spatial hinge $\Pi_{D-2}$, 
one is the $\Pi_{D-1}$ and the other is the 
hyperplane $S$ defined above. The 
ingoing
unit normal to 
$\Pi_{D-1}$ is simply $e_\mathrm{v}$. 
The dihedral angle 
$\phi^{(D-1)\uparrow}$ is then determined by  
\begin{align}
  \cos\phi^{(D-1)\uparrow}=u\cdot e_\mathrm{v}
  =-\frac{\delta L_i}{\sqrt{\delta h_i^2+\delta L_i^2}},  
\end{align}
from which (\ref{eq:phi_down_ptu}) follows.




\section{Derivation of $p(\nu)$}
\label{sec:pnu}
\setcounter{equation}{0}

In this appendix we give a brief account of Eq. (\ref{eq:pnu}). 
As in Sect. \ref{sec:ra} we denote the number of $j$-cubes 
in a $D$-cube by $N^{(D)}_j$. It is given by
\begin{align}
  \label{eq:NDj}
  N^{(D)}_j=2^{D-j}{D \choose j} \qquad (0\leq j\leq D).
\end{align}

Each $j$-dimensional face ($0 \leq j \leq D-1$) of the parent $D$-cube is subdivided into $\nu^j$ $j$-cubes in the geodesic dome $\Gamma_\nu$.
Noting that every three-way connector in $\Gamma_\nu$ comes 
from one of the peaks of the original $D$-cube, we find 
the number of three-way connectors ${\cal N}_3$ in $\Gamma_\nu$ as
\begin{align}
  \label{eq:Ntwc}
  {\cal N}_3=8{D \choose 3}\nu^{D-3}.
\end{align}

Since $\Gamma_\nu$ has $N^{(D)}_{D-1}\nu^{D-1}$ facets and 
each of them contains $N^{(D-1)}_{D-3}$ peaks of $\Gamma_\nu$, ${\cal N}_3$ 
and the number of four-way connectors ${\cal N}_4$ in $\Gamma_\nu$ 
are constrained by 
\begin{align}
  \label{eq:nump}
  3{\cal N}_3+4{\cal N}_4=N^{(D-1)}_{D-3}N^{(D)}_{D-1}\nu^{D-1}. 
\end{align}
This together with (\ref{eq:NDj}) and (\ref{eq:Ntwc}) leads to 
\begin{align}
  {\cal N}_4=6{D \choose 3}(\nu^2-1)\nu^{D-3}.
\end{align}

The averaged number of facets sharing a peak of $\Gamma_\nu$  
is given by 
\begin{align}
  p(\nu)=\frac{3{\cal N}_3+4{\cal N}_4}{{\cal N}_3+{\cal N}_4}.
\end{align}
It yields Eq. (\ref{eq:pnu}).




\end{document}